\documentclass[11pt]{article}

\usepackage[utf8x]{inputenc}

\usepackage{textcomp}

\usepackage{authblk}

\usepackage{setspace}   
\usepackage[right]{lineno}  

\usepackage{subcaption}
\usepackage{wrapfig}
\usepackage{graphicx}
\usepackage{float}
\usepackage{xr-hyper}
\usepackage{hyperref}
\usepackage[super, sort&compress]{natbib}
\usepackage[margin=1in]{geometry}
\usepackage{indentfirst}
\usepackage{orcidlink}
\usepackage[dvipsnames]{xcolor}
\usepackage{soul}
\usepackage{helvet}

\title{Materials Acceleration Platform for Electrochemistry: a Platform for Autonomous Electrochemistry}

\date{}

\author[1,2]{Daniel Persaud\orcidlink{0009-0004-9980-2704}}
\author[1]{Mike Werezak}
\author[1]{Mark Xu}
\author[2]{Melyne Zhou\orcidlink{0009-0009-0498-8101}}
\author[1]{Frank Benkel}
\author[1]{Xin Pang\orcidlink{0009-0007-7722-0991}}
\author[1]{Vahid Attari\orcidlink{0000-0003-3507-8847}}
\author[3]{Brian DeCost\orcidlink{0000-0002-3459-5888}}
\author[2]{Ashley Dale\orcidlink{0000-0001-8233-5258}}
\author[1]{Nicholas Senior\orcidlink{0000-0001-6440-9161}}
\author[1, *]{Gabriel Birsan}
\author[1, 2, 4, 5, 6, **]{Jason Hattrick-Simpers\orcidlink{0000-0003-2937-3188}}

\affil[1]{CanmetMATERIALS, Natural Resources Canada, 183 Longwood Road South, Hamilton, ON, Canada}
\affil[2]{Department of Materials Science and Engineering, University of Toronto, 27 King’s College Cir, Toronto, ON, Canada}
\affil[3]{Material Measurement Laboratory, National Institute of Standards and Technology, 100 Bureau Dr, Gaithersburg, MD, USA}
\affil[4]{Acceleration Consortium, University of Toronto, 27 King’s College Cir, Toronto, ON, Canada.}
\affil[5]{Vector Institute for Artificial Intelligence, 661 University Ave, Toronto, ON, Canada.}
\affil[6]{Schwartz Reisman Institute for Technology and Society, 101 College St, Toronto, ON, Canada.} 
\affil[*]{Correspondence: gabriel.birsan@NRCan-RNCan.gc.ca}
\affil[**]{Correspondence: jason.hattrick.simpers@utoronto.ca}

\begin{document}

\maketitle

\doublespacing    

\begin{abstract}
Corrosion testing is slow, labor-intensive, and sensitive to operator technique, limiting the generation of large, high-quality datasets for data-driven materials discovery. The Materials Acceleration Platform for Electrochemistry (MAP-E) is an autonomous, high-throughput system, capable of performing parallel electrochemical experiments. It integrates robotic liquid handling, sample transfer with a multi-channel potentiostatic control to extract corrosion metrics without human intervention. Validation against an ASTM G61-analog benchmark demonstrates good reproducibility, with a standard deviation of 75 mV in pitting potential across 32 automated measurements. The platform was then employed to autonomously construct pH–chloride stability diagrams for 304 stainless steel using an uncertainty-driven sampling strategy on a Gaussian process surrogate model. This approach reduces operator involvement and accelerates the exploration of environmental spaces. The MAP-E establishes a framework for autonomous electrochemical experimentation, enabling generation of corrosion datasets that inform materials discovery, alloy design, and durability assessment in service environments.
\end{abstract}

\newpage

\section*{Introduction}

The corrosion of metals and alloys remains a major challenge in ensuring the safety and reliability of infrastructure, transportation, and energy systems\cite{Cabana2022,Iannuzzi2022,Xia2022}. During service, materials are exposed to a diverse range of complex environments that can accelerate degradation, leading to significant safety concerns and as a results, substantial economic cost\cite{Groysman2010,Koch2016,Koch2017,Bender2022,Iannuzzi2022,Xia2022, Pence2025}. Predicting how materials will perform under different service conditions, or designing new materials with improved corrosion resistance for specific environments, requires an understanding of the complex interplay between alloy chemistry, microstructure, surface condition, and environmental factors\cite{Nnoka2024,Hariharan2024,Chen2025}. This multifaceted nature makes the systematic exploration of corrosion phenomena experimentally demanding, especially because traditional electrochemical experiments are labor-intensive and sensitive to procedural details, requiring considerable experience to achieve reproducible results\cite{Xia2022,HusseinKhalaf2024}. As a result, generating large, high-quality datasets that capture corrosion behavior across broad parameter spaces is challenging, limiting the ability to establish predictive relationships between material properties, environment, and degradation mechanisms.

Autonomous electrochemical platforms have the potential to address these challenges by integrating networked instrumentation, modular fluid handling, and parallelized electrochemical cells under unified control software\cite{Stier2024,Pence2025,Nwabara2025}. When coupled with machine learning (ML), these systems can be used to generate informative datasets through efficient exploration of complex experimental spaces to better understand underlying patterns. The datasets produced can be shared with the broader research community, enabling others to develop predictive models, identify promising research directions, and collaborate to accelerate innovation. 

Recent years have seen substantial progress in high-throughput (HT) electrochemical platforms, including systems that integrate hardware and software to varying degrees of autonomous operation. Early HT approaches emphasized parallel screening to maximize throughput, often relying on shared electrolytes, reduced solution volumes, simplified electrode configurations, or measurements limited to post-exposure or optical characterization\cite{Pimenova2008,Taylor2011,White2012,Shi2016,Gerroll2023}. More recent efforts have shifted toward increased autonomy through integrated hardware–software platforms, commonly based on scanning droplet cells or gantry-based systems that sequentially position a shared reference and counter electrode into electrolyte-filled wells to characterize the sample below\cite{Joress2022,DeCost2022,Oh2023,Rial-Rodriguez2024,Jenewein2024,Quinn2024,Sheng2024,Abed2024,Darvish2025,Fisker-Bodker2025}. Despite these advances, most platforms face trade-offs: systems closer to autonomy often perform experiments serially, limiting speed, whereas parallelized systems frequently compromise on experimental fidelity, restricting their relevance for corrosion science. These accommodations are typically driven by constraints in hardware, cost, or the complexity associated with linking multiple pieces of equipment\cite{Christensen2021,Adam2024}. Collectively, these limitations define a critical gap for corrosion research: the absence of a platform that can perform parallel electrochemical measurements, using traditional cell geometries, electrolyte volumes, and standard electrode configurations, features necessary for high-quality data-driven exploration across diverse material-environment combinations.

To address this gap, we have developed the Materials Acceleration Platform for Electrochemistry (MAP-E), a flexible, HT platform designed for autonomous electrochemical experiments. It integrates eight individually addressable electrochemical flat-cells, each capable of performing independent measurements in parallel. The platform features modular fluid handling to explore a wide range of aqueous environments. Robust component control and scheduling software manages complex experimental campaigns, enabling automated operation. With these capabilities, the MAP-E can generate the precise and reliable electrochemical data required for applying ML algorithms to autonomously explore experimental spaces.

In this work, we first validate the MAP-E’s ability to perform reproducible electrochemical measurements using a modified protocol derived from an established corrosion standard. We then demonstrate its ability to autonomously construct data-driven stability diagrams within a single closed-loop campaign, demonstrating a more efficient use of operator time compared to traditional sequential approaches. Once the stock solutions are prepared and samples are loaded, the system loads samples and mixes solution to explore the defined experimental space without further human intervention. By running eight cells in parallel, the MAP-E achieves an immediate eightfold increase in throughput. An uncertainty-driven acquisition strategy is used to adaptively select subsequent experimental conditions based on the model’s posterior predictive uncertainty. This focuses sampling on regions where the model is least certain, allowing informative areas of the design space to be explored efficiently under a constrained experimental budget. By producing standard-aligned, reproducible data and demonstrating the autonomous exploration of experimental spaces, the MAP-E can support collaborative, data-driven materials discovery through the generation and public sharing of high-fidelity experimental datasets. 

\section*{Results and Discussion}
\label{section:resultsAndDiscussion}

\subsection*{Corrosion Validation}
\label{section:corrosionValidation}
As discussed above, the design and configuration of electrochemical platforms vary widely, making comparison across laboratories difficult\cite{Nwabara2025}, and with eight independent cells, it is also necessary to verify intra-platform reproducibility\cite{Kulesa2024}. Accordingly, we aimed to validate the electrochemical performance of the MAP-E using experiments aligned with standards commonly employed in corrosion testing to assess reproducibility. These experiments were performed to demonstrate that the MAP-E can generate reliable electrochemical data suitable for subsequent adaptive experimentation.

To validate the electrochemical performance of the MAP-E, we performed potentiodynamic polarization (PP) experiments using an analog to ASTM G61\cite{ASTMG61}, a standard for evaluating localized corrosion susceptibility of alloys in aqueous chloride (Cl$^-$) environments. Following as closely as possible, the procedure allows us to quantitatively benchmark the MAP-E’s performance against established metrics for $E_{\mathrm{pit}}$. ASTM G61 defines expected inter-laboratory reproducibility and bias in $E_{\mathrm{pit}}$ during PP measurements. Based on results from five independent laboratories, the standard reports an inter-laboratory 2$\sigma$ of $\approx$ 600 mV in $E_{\mathrm{pit}}$. However, it does not define intra-laboratory repeatability, leaving the expected variability when multiple identical tests are performed within a single laboratory, or in the case of the MAP-E, a single cell, unspecified.

Several adaptations to ASTM G61 were made to accommodate the MAP-E's capabilities. The platform operates under ambient (aerated) conditions rather than the nitrogen-deaerated environment specified by ASTM G61. An Ag/AgCl (1 mol/L KCl) reference electrode (RE) was used in place of the saturated calomel electrode due to its compatibility with the MAP-E’s automated handling system, lower maintenance requirements, and reduced toxicity, while still providing stable performance in Cl$^-$-containing electrolytes. Additionally, the specimen holder was implemented using the MAP-E’s custom flat-cell design, and tests were performed at room temperature, as opposed to the 25 \textdegree C specified in the standard. While these modifications reflect practical aspects of operating an autonomous, multi-cell platform, the essential elements of the ASTM G61 methodology; electrolyte composition, electrochemical parameters, and interpretation of $E_{\mathrm{pit}}$, are preserved, enabling a meaningful assessment of the MAP-E’s electrochemical capability.

\begin{figure}[H]
    \centering
    \textbf{Results from the ASTM G61 replication experiments with the MAP-E}\par\medskip
    \includegraphics[width=0.95\textwidth]{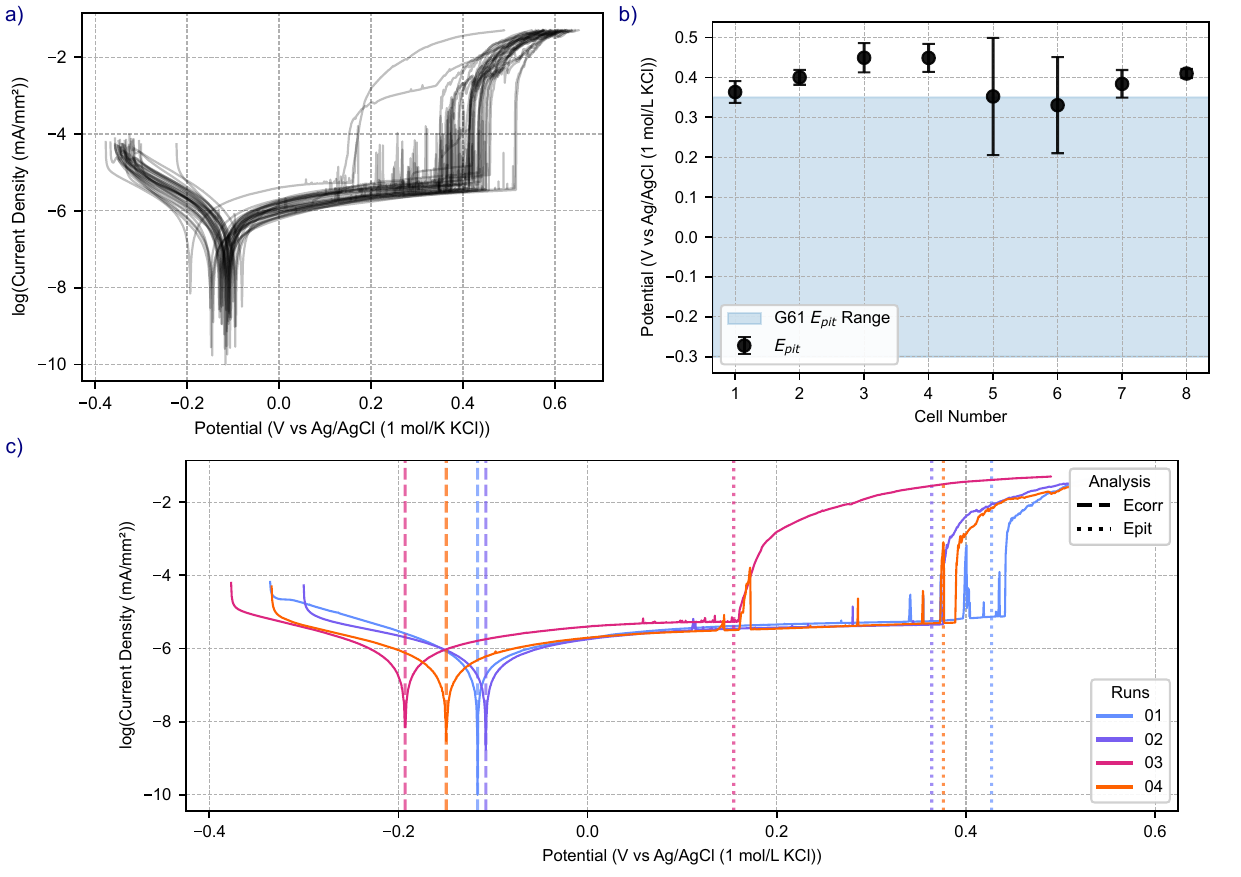}
    \caption{a) Potentiodynamic polarization curves for the all 304 stainless steel samples (32) in aerated aqueous NaCl solution with mass fraction of 3.56 wt. ~\% at room temperature, measured across the eight cells of the MAP-E. b) Distribution of $E_{\mathrm{pit}}$ extracted from the polarization curves, showing cell-to-cell variability and overall reproducibility. Error bars indicate spread across measurements within each cell. c) Polarization curves from just one cell (Cell 6) to illustrate run-to-run stochasticity within a single cell.}
    \label{figure:corrosionValidation}
\end{figure}

All eight REs are calibrated against a common standard before use, and any drift from the standard is corrected for in the data analysis to ensure the results reflect electrochemical differences rather than RE inconsistencies. \hyperref[figure:corrosionValidation]{Figure 1a} shows the results from the ASTM G61-analog validation experiments. After a 1 hour rest, PP was performed in aerated 3.56 wt. \% NaCl solution at room temperature, at a scan rate of 0.167 mV s$^{-1}$, from -250 mV vs. $E_{corr}$ until 5 mA cm$^{-2}$ was reached. Thirty-two 304 stainless steel samples were prepared according to the ASTM procedure and tested across four full the MAP-E runs (8 unique samples per run), rinsing the cell twice with type 1 water between runs to rinse cells and the REs. E$_{\mathrm{pit}}$ was extracted using the automated data analysis pipeline described in the \hyperref[section:adaptiveExperimentationFramework]{adaptive experimentation framework section}.

\vspace{12pt}

Across all measurements (PP curves available in \hyperref[si:corrosionValidation]{supplementary information section S1}) the mean $E_{\mathrm{corr}}$ was -119 mV ± 21 mV and the mean $E_{\mathrm{pit}}$ was 392 mV ± 75 mV. The standard deviation of $E_{\mathrm{pit}}$ across the MAP-E measurements is approximately four times smaller than the inter-laboratory standard deviation implied by ASTM G61 ($\approx$300 mV, based on a reported 2$\sigma$ spread of $\approx$600 mV across multiple laboratories). At the cell level, variability was even tighter, with standard deviations in most cells $<$40 mV, shown in \hyperref[figure:corrosionValidation]{Figure 1b}, demonstrating consistent performance in repeated tests within individual cells. The $E_{\mathrm{pit}}$ values obtained on the MAP-E were consistently higher than those reported in ASTM G61 datasets, with individual measurements up to $\approx$100 mV higher and a mean positive offset of $\approx$40 mV.

The MAP-E results demonstrate high reproducibility and precision, with inter-cell variability below the inter-laboratory spread defined in ASTM G61. The elevated average $E_{\mathrm{pit}}$ and the measurement noise in the passive region are both attributable to testing in aerated electrolyte. Dissolved oxygen increases cathodic reaction kinetics and promotes passive film stabilization, which shifts the pitting potential positively relative to deaerated ASTM conditions\cite{Zeng2021}. The presence of dissolved oxygen also introduces metastable pitting events that manifest as fluctuations in the current response near $E_{\mathrm{pit}}$.\cite{Pistorius1992,Pistorius1994}. This behavior is illustrated in \hyperref[figure:corrosionValidation]{Figure 1c}, where early pit initiation is observed during two nominally identical runs (3 and 4): in one case the metastable pit transitioned into stable pit growth, whereas in the other it repassivated at approximately the same potential. Notably, these ‘early-pit’ events occur near the mean $E_{\mathrm{pit}}$ reported in ASTM G61, further supporting the interpretation that the observed differences arise from environmental aeration. 

In addition to producing measurement quality consistent with established corrosion test standards, the MAP-E executed all steps of the experiment automatically once samples and stock solutions are setup. In this validation study, eight cells were operated in parallel, enabling all the potentiodynamic tests to be completed with minimal operator involvement beyond routine monitoring over the course of two days. Although the electrochemical runtime remains comparable to conventional testing, operator time was reduced substantially. This capability positions the MAP-E to not only accelerate data generation, but also to support quantitative, data-drive studies.

\subsection*{Autonomous Stability Diagram}

Having established the MAP-E’s reproducibility and precision, we deployed the platform to autonomously generate empirical corrosion stability diagrams. Traditional stability diagrams, such as Pourbaix diagrams, describe the thermodynamic stability of metals\cite{Delahay1950}, while experimentally generated diagrams relating applied potential to pitting susceptibility, such as Pedeferri diagrams\cite{Bertolini2009, Pedeferri1996}, are typically constructed manually and limited to a small number of discrete conditions. The MAP-E extends these concepts by performing PP experiments across pH-Cl$^-$ space and selecting subsequent conditions based on a model's predictive uncertainty. A Gaussian process (GP) surrogate model predicts the $E_{\mathrm{pit}}$ throughout the design space, and an uncertainty-driven acquisition function guides the platform toward the most informative regions, reducing redundant experiments, as described in the \hyperref[section:adaptiveExperimentationFramework]{adaptive experimentation framework section}. While demonstrated here for 304 stainless steel, this workflow is generalizable, enabling data-driven, autonomous mapping of corrosion susceptibility and understanding of the service envelope of materials across complex environmental conditions.

The exploration space for autonomous corrosion mapping was defined in terms of pH and Cl$^-$, two critical variables controlling corrosion behavior for stainless steels and following the concept of Pedeferri diagrams\cite{Bertolini2009, Pedeferri1996}. The MAP-E investigated environments spanning pH 3  to 10.5 (0.5-unit increments) and Cl$^-$ from 0 mol/L to 0.1 mol/L (semi-logarithmic increments), encompassing 80 unique conditions. Each environment was formulated using a BR buffer as the background electrolyte to maintain pH stability while covering a broad range of pH values from a limited set of stock solutions, with controlled additions of sodium chloride stocks achieving the target Cl$^-$. A Python-based framework calculates the precise stock volumes to reach the target pH and Cl$^-$, after which the liquid handling system automatically prepares each condition and measures the pH in the mixing tank. We observed that the measured pH differ slightly from the target value due to the pumping accuracy but the pH is always measured in the mixing tank prior to dispensing and the measured value that is used for subsequent modeling. Each condition was then characterized electrochemically using the same procedure as the \hyperref[section:corrosionValidation]{validation measurements}, with the automated data extraction workflow described in the \hyperref[section:adaptiveExperimentationFramework]{adaptive experimentation framework section}, generating the $E_{\mathrm{pit}}$ for the subsequent modeling.

\newpage
\begin{figure}[H]
    \centering
    \textbf{Workflow diagram of the autonomous experimental campaign}\par\medskip
    \includegraphics[width=\textwidth]{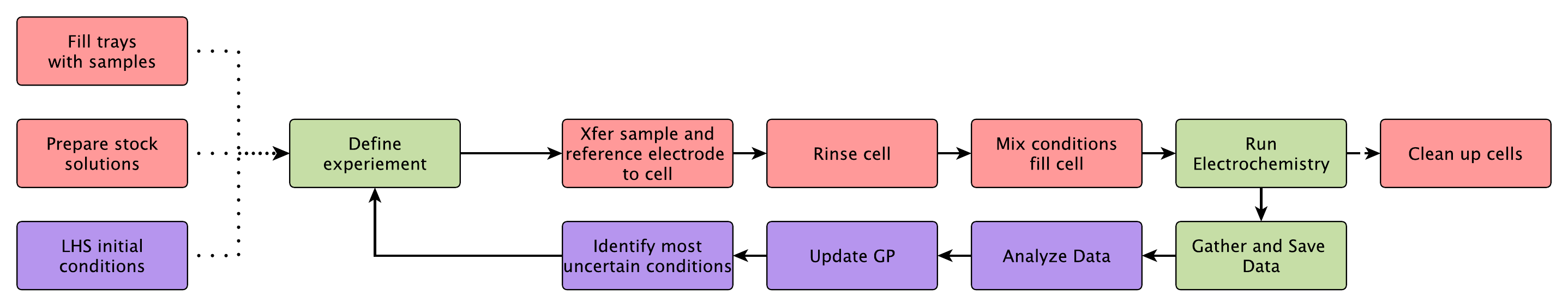}
    \caption{High-level workflow diagram of the autonomous experimental workflow for constructing pH–Cl$^-$ stability diagrams using the MAP-E platform. The red boxes indicate steps that can primarily be associated with the hardware, the green boxes indicate software-driven steps, and the purple boxes indicate the adaptive modeling components, see \hyperref{section:methods}[methods section] for details.}
    \label{figure:autonomousWorkflow}
\end{figure}

The autonomous experimental workflow is illustrated in \hyperref[figure:autonomousWorkflow]{Figure 2}. The campaign was initialized with four conditions selected using Latin hypercube sampling (LHS), with duplicate measurements for each condition to capture and model both intrinsic stochasticity in experiments observed in the \hyperref[section:corrosionValidation]{validation section}, and any misidentification of $E_{\mathrm{pit}}$ from the PP analysis pipeline, discussed in \hyperref[si:autonomousStabilityDiagramPPScans]{supplementary information section S2}. After the initial measurements, the MAP-E proceeded with a fixed budget of 48 additional samples (three trays, 24 conditions in duplicate), autonomously selecting subsequent conditions according to the uncertainty-driven sampling strategy described in the \hyperref[section:adaptiveExperimentationFramework]{adaptive experimentation framework section}. These steps provide a systematic method of exploring environmental space to construct empirical stability diagrams.

\begin{figure}[H]
    \centering
    \textbf{Results from the autonomous stability diagram campaign}\par\medskip
    \includegraphics[width=0.85\textwidth]{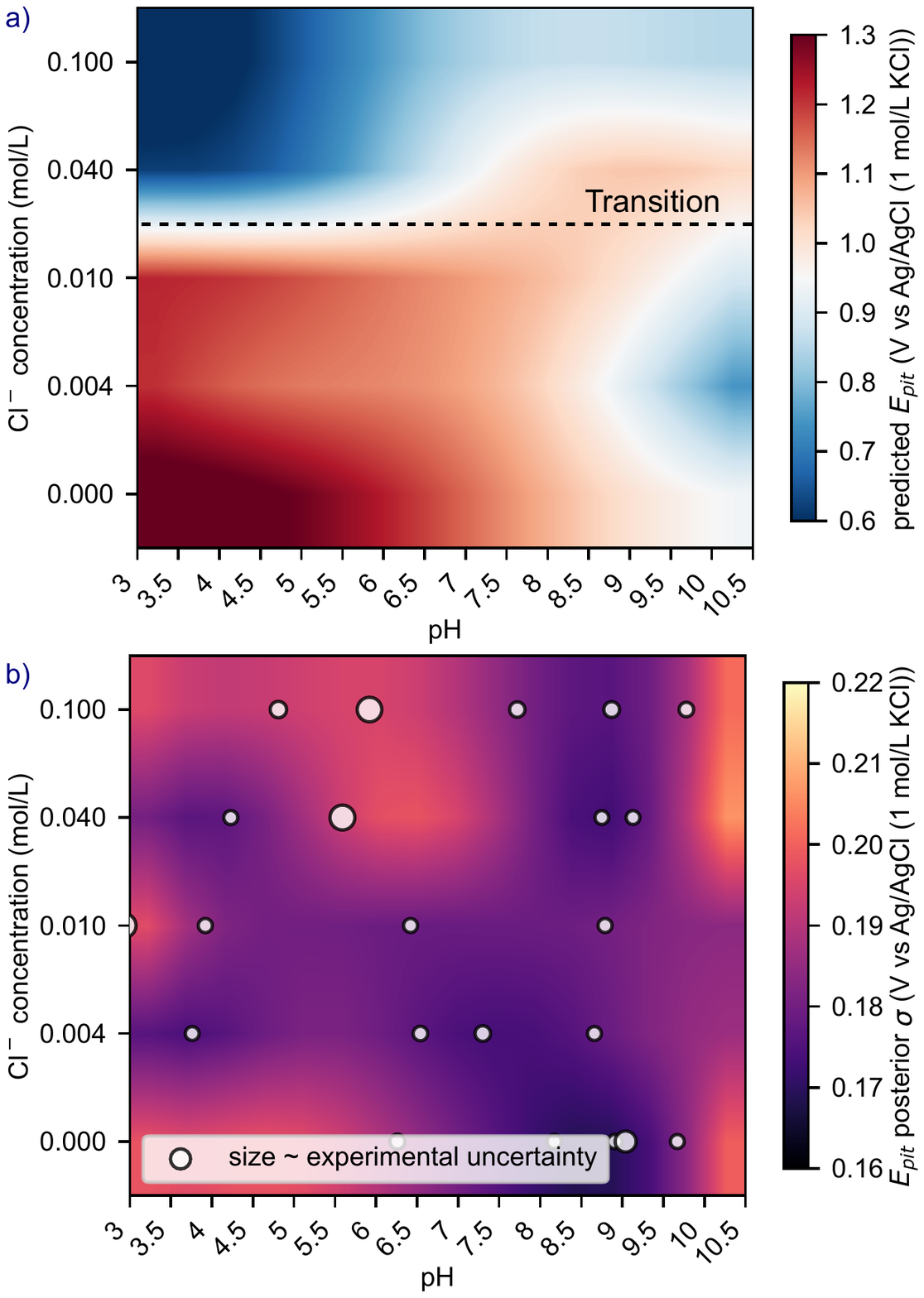}
    \caption{a) Stability diagram showing the predicted $E_{\mathrm{pit}}$ across the pH-Cl$^-$ space for 304 stainless steel, generated autonomously by the MAP-E. b) Uncertainty map corresponding to the stability diagram, highlighting regions where the model is less certain about $E_{\mathrm{pit}}$ predictions.}
    \label{figure:stabilityDiagramResults}
\end{figure}

The resulting stability diagram, \hyperref[figure:stabilityDiagramResults]{Figure 3a}, shows that 304 stainless steel is least susceptible to pitting in oxygenated, moderately acidic, low-Cl$^-$ environments and most susceptible under high-Cl$^-$, acidic conditions. When compared with other conventional mapping studies on 304 stainless steel, the MAP-E-generated stability diagram reproduces the expected decrease in $E_{\mathrm{pit}}$ with increasing Cl$^-$, consistent with increased susceptibility to localized corrosion\cite{Belle2017,Dastgerdi2019,Dastgerdi2019a}. However, a transition region appears, above which $E_{\mathrm{pit}}$ decreases with decreasing pH, whereas below this region, the opposite pH dependence is observed. Additionally, all $E_{\mathrm{pit}}$ values observed in this study are higher ($\approx$400mV) than those reported in similar mapping studies investigating pitting susceptibility of stainless steel in Cl$^-$ environments using conventional solutions\cite{Belle2017,Dastgerdi2019,Dastgerdi2019a,Wang2019,Wetzel2022}. The elevated $E_{\mathrm{pit}}$ values are consistent with the phosphate and borate components of the BR buffer, which have been shown to enhance passive film stability and increase resistance to pitting \cite{Wu2018,Hashimoto2026}. These buffering species may also contribute to the observed transition in pH dependence at low Cl$^-$ concentrations.

The GP-derived stability diagram resolves this change in behavior across the pH-Cl$^-$ space without embedding mechanistic corrosion relationships. At high Cl$^-$ concentrations, the observed trends are consistent with established understanding: sufficient Cl$^-$ promotes pit initiation, while decreasing pH further destabilizes the passive film and lowers the potential required for pit propagation\cite{Kolotyrkin1963,Flint1980}. In these conditions, aggressive Cl$^-$ adsorption and reduced oxide stability dominate the pitting response. Below the transition region, Cl$^-$ concentrations appear insufficient for Cl$^-$-driven pitting processes to dominate the electrochemical response. In this area, the observed variation in $E_{\mathrm{pit}}$ is likely influenced by the phosphate- and borate-containing buffer chemistry, which promotes passive film stability, and is consistent with published measurements of stainless steel pitting behavior in phosphate- and borate-containing Cl$^-$ electrolytes\cite{Salvarezza1987,Lakatos-Varsanyi1998,Wu2018,Hashimoto2026}.

The posterior predictive uncertainty map, \hyperref[figure:stabilityDiagramResults]{Figure 3b}, highlights regions where additional data would most improve model confidence, with higher uncertainty near the boundaries of the design space where experimental coverage is sparse and where measurements show more variability. Despite these localized regions, the uncertainty surface remains relatively uniform, with an average predictive uncertainty of $\approx$175 mV. Together, these results define the service envelope of 304 stainless steel across the investigated pH–Cl$^-$ space and demonstrate that the MAP-E platform can guide autonomous experimental campaigns.

In this work we present the MAP-E, a modular, HT system capable of autonomously performing parallel electrochemical experiments with individually addressable cells. Validation against a ASTM G61-analog benchmark confirms reproducibility and quantitative precision, with standard deviations in $E_{\mathrm{pit}}$ approximately one quarter of the inter-laboratory variability reported in the standard, producing data suitable for autonomous studies. Using a posterior-uncertainty–driven experimental framework, the MAP-E autonomously generated empirical pH–Cl$^-$ stability diagrams for 304 stainless steel, capturing trends in pitting susceptibility and associated uncertainties. The platform’s architecture enables closed-loop experimentation: selecting, executing, and analyzing experiments without human intervention, while supporting flexible definition of electrochemical workflows beyond corrosion testing. Collectively, these results establish the MAP-E as a generalizable and data-driven framework for mapping corrosion service envelopes, and demonstrate its potential to accelerate both corrosion research and broader electrochemical discovery efforts.

\section*{Methods}
\label{section:methods}

The design philosophy of the MAP-E centers on maximizing experimental throughput while minimizing the trade-offs in versatility and flexibility that commonly constrain automated electrochemical platforms. Prior to operation, stock solutions are prepared to span the target environmental range, and polished samples are loaded into the dedicated sample racks. A high-level software interface accepts experiment requests which specify the electrochemical methods, sample selection, and environmental conditions. Once initiated, the MAP-E automatically orchestrates the full workflow. The system loads samples into the electrochemical cells, mixes and dispenses the target electrolyte compositions via the liquid-handling system. It executes the prescribed electrochemical measurements, and processes the resulting data to extract the corrosion parameters of interest while the system rinses the cells and prepares for subsequent experiments. The following sections describe the hardware, control software, and modeling framework for the autonomous stability diagram investigation.

\subsection*{Hardware}
\label{section:hardware}

The MAP-E hardware is organized into four subsystems that enable automated, HT electrochemical experiments: (1) the electrochemical cell array, (2) gantry system, (3) liquid handling, and (4) the multi-channel potentiostat.

\begin{figure}[H]
    \centering
    \textbf{Rendering of the key hardware subsystems of the MAP-E platform.}\par\medskip
    \includegraphics[width=\textwidth]{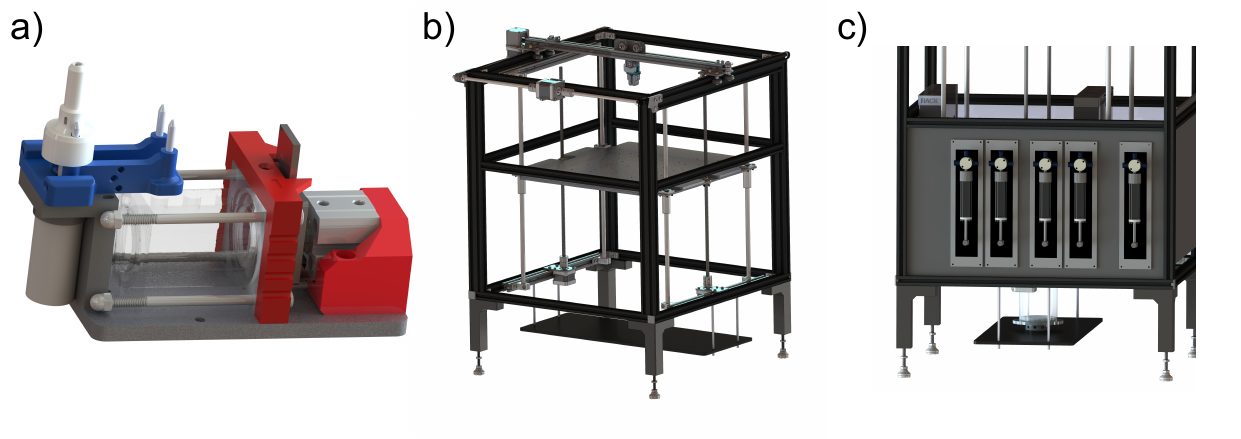}
    \caption{a) Custom electrochemical flat cell with pneumatic clamping mechanism and integrated electrodes (reference electrode in storage tube on the left, counter electrode mesh in the rear end cap (not visible), and sample as the working electrode clamped by the pneumatic cylinder on the right). b) Gantry system adapted from a 3D printer. c) One side of the liquid handling system, showing 5 of the 10 dosing pumps and the central mixing tank attached to the gantry frame, The remaining pumps are mounted on the opposite side of the gantry and are not visible in this rendering.}
    \label{figure:hardware}
\end{figure}

The electrochemical array consists of eight independent, custom flat cells, constructed from a cylindrical polycarbonate (PC) body with two machined PC end caps. Each cell contains a platinum mesh counter electrode (CE) in the rear end cap, while the front end cap has a 1 cm$^2$ hole sealed with a Viton o-ring. A pneumatic cylinder mounted on the front end cap, when triggered, secures a sample against the o-ring. This maintains a 25 mL electrolyte volume that meets the 0.2 mL/mm$^2$ minimum ratio recommended by ASTM G31\cite{ASTMG31}. When not in use, the CH Instruments Ag/AgCl (1 mol/L KCl) REs are stored in a tube mounted outside the cell, shown in \hyperref[figure:hardware]{Figure 4a}.

A retrofitted 3D printer serves as the Cartesian gantry system and structural backbone for the platform, shown in \hyperref[figure:hardware]{Figure 4b}. The gantry's platform supports the two sample racks, each accommodating 16 samples (60 mm × 20 mm × 3 mm), and the electrochemical cell array. The gantry is responsible for transferring samples from the racks, as well as moving the REs from their storage tubes into the cells. 

Liquid handling is performed by a network of ten 25 mL Tricontinent CX6000 dosing pumps with multi-port valves mounted on the gantry frame, shown in \hyperref[figure:hardware]{Figure 4c}, all connected to a central mixing tank under the gantry equipped with a pH meter and magnetic stir bar. Eight of these pumps are dedicated to delivering electrolyte from the mixing tank to the cells and removing used solution to waste, while the remaining two are used to fill the mixing tank with stock and cleaning solutions, the latter of which can be used to flush cells between experiments.

A Bio-Logic VMP-3 multi-channel potentiostat provides independent electrical control of the eight cells. Connections from the potentiostat are routed through a custom breakout box with gold-plated connectors, ensuring reliable contact. For each cell the electrical connections are made from the breakout box to: the RE through a 2mm banana connector in a 3D-printed holder that enables the gantry to move the RE into position, and easy maintenance; the CE via a fixed ring terminal connecting to a wire attached to the platinum mesh; and the working electrode through a conductive plate at the end of the pneumatic cylinder, which engages the sample upon clamping. 

\subsection*{Control Software}
\begin{wrapfigure}{r}{0.45\textwidth}   
    \centering
    \includegraphics[width=0.3\textwidth]{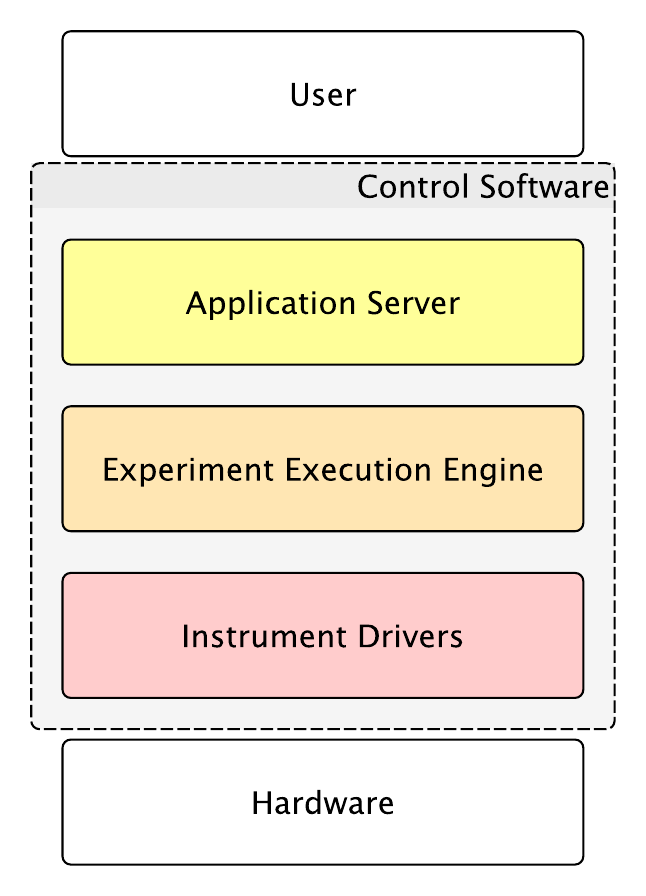}
    \caption{Schematic of the MAP-E software architecture, illustrating the three primary layers: application server, experiment execution engine, and instrument driver libraries.}
    \label{figure:softwareArchitecture}
\end{wrapfigure}

The MAP-E control software was developed as a modular framework for automated, concurrent experimentation. The architecture comprises three layers: (1) the application server, (2) the experiment execution engine, and (3) the instrument driver libraries, shown in \hyperref[figure:softwareArchitecture]{Figure 5}. Together, these layers abstract the underlying hardware complexity and provide a high-level user interface for submitting experiment requests.

\newpage

\begin{figure}[H]
    \centering
    \includegraphics[width=\textwidth]{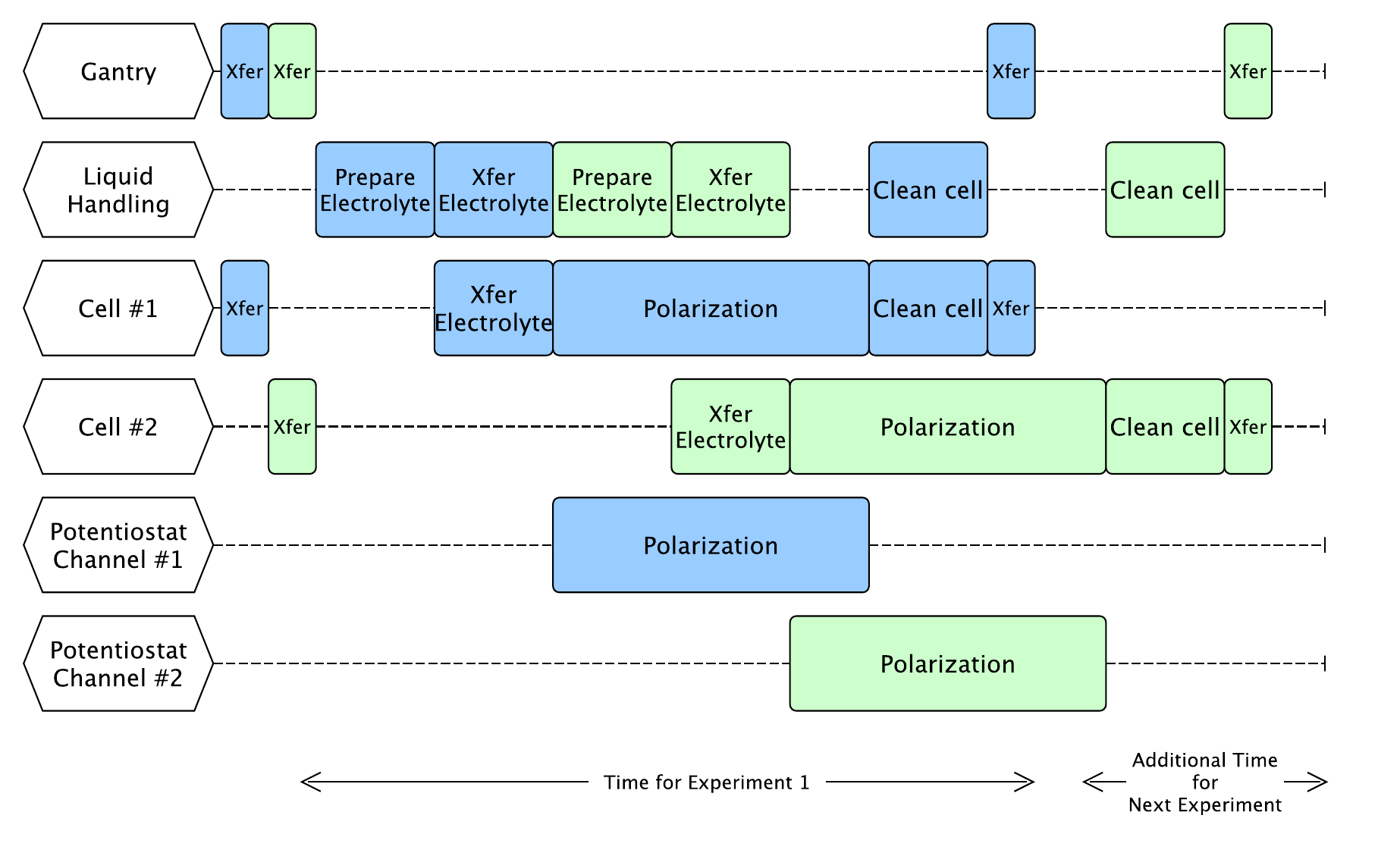}
    \caption{Illustration of the resource allocation and scheduling framework within the experiment execution engine, demonstrating parallel operation of multiple electrochemical cells while managing shared resources like the mixing tank. In this example, separate experiments are denoted by different colors and only two cells are shown for clarity.}
    \label{figure:resourceAllocation}
\end{figure}

The application server provides a network-accessible interface for managing experiment requests and returning the resulting measurement data. The server accepts experiment specifications as structured inputs, handles data collection and storage automatically, as well as maintaining a registry of active and completed jobs to support asynchronous submission, execution, and monitoring. At the core of the control architecture is the experiment execution engine, which orchestrates real-time operation of the hardware subsystems. Experimental procedures are defined as sequences of schedulable tasks (e.g., sample transfer, cell filling, cell cleaning, electrochemical techniques) that are dynamically allocated to available cells or shared resources (e.g., the mixing tank), as illustrated in \hyperref[figure:resourceAllocation]{Figure 6}. This architecture supports fully concurrent operation, allowing multiple experiments to proceed in parallel, while managing shared hardware dependencies transparently. The instrument driver libraries form the interface layer between the execution engine and the physical devices. Each driver exposes the essential functions of its corresponding hardware—such as pumps, actuators, or potentiostat channels—through a consistent software interface, enabling coordination within the broader system. Its modular design allows new instruments or sensors to be incorporated with minimal software modification. 

\subsection*{Adaptive Experimentation Framework}
\label{section:adaptiveExperimentationFramework}

In this work, we perform PP measurements and the resulting data is retrieved from the control software to a user-specified directory. Data is processed automatically, and new experiment requests can then be generated and submitted back to the control software. The raw PP data is analyzed using a Python-based pipeline that identifies the corrosion potential ($E_{\mathrm{corr}}$) and pitting potential ($E_{\mathrm{pit}}$). To mitigate experimental noise, the logarithm of the current density is smoothed using a Savitzky–Golay filter (second-order polynomial) with respect to potential. To further stabilize the second derivative signal, a rolling mean is applied to the resulting derivative. The $E_{\mathrm{corr}}$ is identified as the potential corresponding to the minimum absolute current. The $E_{\mathrm{pit}}$ is determined by locating the maximum of the second derivative ($\frac{d^2\log(i)}{dE^2}$) outside of the $E_{\mathrm{corr}}$ area, corresponding to the onset of stable pitting. These parameters are stored and made available for subsequent modeling.

A GP regression model serves as the surrogate model for predicting $E_{\mathrm{pit}}$ across the design space. The GP employs a scaled Matérn covariance kernel ($\nu = 2.5$) with automatic relevance determination (ARD) to capture varying correlation length scales in each input dimension. A heteroscedastic likelihood is used to represent measurement-dependent noise, accounting for variability between duplicate experiments. The model is initialized with four Latin hypercube–sampled (LHS) conditions and iteratively updated as new data are collected. Experimental selection is guided by an uncertainty-driven acquisition function that prioritizes conditions with the highest posterior predictive variance in $E_{\mathrm{pit}}$. This approach directs the MAP-E toward regions of high model uncertainty or experimental disagreement, balancing exploration of untested conditions with refinement of noisy regions.  

\subsection*{Materials and Sample Preparation}
Standard Type 304 stainless steel (UNS S30403) sheet stock was procured from McMaster-Carr, water-jet cut to size, and polished to 600-grit following ASTM G61. For the validation study, a 3.56 wt. \% sodium chloride solution was prepared according to ASTM G61\cite{ASTMG61}. For the autonomous stability diagram experiments, stock solutions for the Britton-Robinson (BR)\cite{Mongay1974} buffer were prepared using reagent-grade chemicals at the following concentrations: sodium hydroxide (0.3 mol/L), phosphoric acid (0.2 mol/L), acetic acid (0.2 mol/L), and boric acid (0.2 mol/L). Two sodium chloride stock solutions were also prepared to span the Cl$^-$ range: 0.04 mol/L and 0.4 mol/L. All solutions were made with type 1 water (resistivity $>$18  mol/L$\Omega \cdot$ cm$^2$ at 25 \textdegree C) and mixed to generate the desired electrolyte compositions for their respective experiments.

\newpage
\section*{Data Availability}
\label{section:dataAvailability}
The data generated by the MAP-E during the validation and autonomous stability diagram experiments, including raw electrochemical measurements, processed corrosion parameters, and the resulting stability diagrams, are available at the following \href{https://github.com/dpersaud/MAP_E}{GitHub repository} : github.com/dper- saud/MAP\_E.

\section*{Code Availability}
\label{section:codeAvailability}
The code the to plot the data, the GP surrogate modeling, and the adaptive experimental framework is available at the following \href{https://github.com/dpersaud/MAP_E}{GitHub repository} : github.com/dpersaud/MAP\_E.

\section*{Acknowledgements}

We acknowledgement funding from Natural Sciences and Engineering Research Council of Canada, grant \#RGPIN-2023-04843. The research was also, in part, made possible thanks to funding provided to the University of Toronto’s Acceleration Consortium by the Canada First Research Excellence Fund (CFREF-2022-00042). Certain equipment, instruments, software, or materials are identified in this paper in order to specify the experimental procedure adequately, Such identification is not intended to imply recommendation or endorsement of any product or service by NIST, nor is it intended to imply that the materials or equipment identified are necessarily the best available for the purpose.

\newpage
\section*{Author Contributions}

Contributions are defined here after CRediT (the Contributor Roles Taxonomy). D.P: Conceptualization, Methodology, Software, Validation, Formal analysis, Investigation, Data Curation, Writing - Original Draft, Writing - Review \& Editing, Visualization, Project administration. M.W: Methodology, Software, Validation, Writing - Original Draft, Writing - Review \& Editing. M.X: Methodology, Software, Validation, Resources, Writing - Review \& Editing. M.Z: Software. F.B: Resources. X.P: Supervision, V.A: Supervision. B.D: Supervision. A.D: Supervision. N.S: Supervision. G.B: Methodology, Resources, Writing - Original Draft, Funding acquisition. J.H.S: Conceptualization, Writing - Review \& Editing, Supervision, Funding acquisition.

\section*{Competing Interests}

The authors declare no competing financial or non-financial interests.

\newpage
\bibliographystyle{naturemag}
\bibliography{lib}

@article{Abed2024,
  title = {{{AMPERE}}: Automated Modular Platform for Expedited and Reproducible Electrochemical Testing},
  shorttitle = {{{AMPERE}}},
  author = {Abed, Jehad and Bai, Yang and Persaud, Daniel and Kim, Jiheon and Witt, Julia and {Hattrick-Simpers}, Jason and Sargent, Edward H.},
  year = 2024,
  journal = {Digital Discovery},
  volume = {3},
  number = {11},
  pages = {2265--2274},
  issn = {2635-098X},
  doi = {10.1039/D4DD00203B},
  urldate = {2025-04-15},
  langid = {english}
}

@article{Adam2024,
  title = {The Automated Lab of Tomorrow},
  author = {Adam, David},
  year = 2024,
  month = apr,
  journal = {Proc. Natl. Acad. Sci. U.S.A.},
  volume = {121},
  number = {17},
  pages = {e2406320121},
  issn = {0027-8424, 1091-6490},
  doi = {10.1073/pnas.2406320121},
  urldate = {2025-10-09},
  langid = {english}
}

@misc{ASTMG31,
  title = {Standard {{Practice}} for {{Laboratory Immersion Corrosion Testing}} of {{Metals}}},
  year = 1999,
  number = {G31}
}

@misc{ASTMG61,
  title = {Test {{Method}} for {{Conducting Cyclic Potentiodynamic Polarization Measurements}} for {{Localized Corrosion Susceptibility}} of {{Iron-}}, {{Nickel-}}, or {{Cobalt-Based Alloys}}},
  author = {{G01 Committee}},
  publisher = {ASTM International},
  doi = {10.1520/G0061-86R18},
  urldate = {2025-04-15},
  langid = {english}
}

@phdthesis{Belle2017,
  type = {{{MASc}}},
  title = {Pedeferri {{Diagrams}} of {{Stainless Steel}}: {{Study}} of the {{Effect}} of {{Temperature}} and {{pH}} by Means of {{Design}} of {{Experiment}}},
  author = {Belle, Umberto},
  year = 2017,
  school = {POLITECNICO DI MILANO}
}

@article{Bender2022,
  title = {Corrosion Challenges towards a Sustainable Society},
  author = {Bender, Roman and F{\'e}ron, Damien and Mills, Douglas and Ritter, Stefan and B{\"a}{\ss}ler, Ralph and Bettge, Dirk and De Graeve, Iris and Dugstad, Arne and Grassini, Sabrina and Hack, Theo and Halama, Maros and Han, En-Hou and Harder, Thomas and Hinds, Gareth and Kittel, Jean and Krieg, Romina and Leygraf, Christofer and Martinelli, Laure and Mol, Arjan and Neff, Delphine and Nilsson, Jan-Olov and Odnevall, Inger and Paterson, Steve and Paul, Shiladitya and Pro{\v s}ek, Tom{\'a}{\v s} and Raupach, Michael and Revilla, Reynier I. and Ropital, Fran{\c c}ois and Schweigart, Helmut and Szala, Elizabeth and Terryn, Herman and Tidblad, Johan and Virtanen, Sannakaisa and Volovitch, Polina and Watkinson, David and Wilms, Marc and Winning, George and Zheludkevich, Mikhail},
  year = 2022,
  month = nov,
  journal = {Materials \& Corrosion},
  volume = {73},
  number = {11},
  pages = {1730--1751},
  issn = {0947-5117, 1521-4176},
  doi = {10.1002/maco.202213140},
  urldate = {2023-12-12},
  langid = {english}
}

@article{Bertolini2009,
  title = {Effects of Cathodic Prevention on the Chloride Threshold for Steel Corrosion in Concrete},
  author = {Bertolini, L. and Bolzoni, F. and Gastaldi, M. and Pastore, T. and Pedeferri, P. and Redaelli, E.},
  year = 2009,
  month = feb,
  journal = {Electrochimica Acta},
  volume = {54},
  number = {5},
  pages = {1452--1463},
  issn = {00134686},
  doi = {10.1016/j.electacta.2008.09.033},
  urldate = {2025-10-15},
  copyright = {https://www.elsevier.com/tdm/userlicense/1.0/},
  langid = {english}
}

@article{Cabana2022,
  title = {{{NGenE}} 2021: {{Electrochemistry Is Everywhere}}},
  shorttitle = {{{NGenE}} 2021},
  author = {Cabana, Jordi and Alaan, Thomas and Crabtree, George W. and Hatzell, Marta C. and Manthiram, Karthish and Steingart, Daniel A. and Zenyuk, Iryna and Jiao, Feng and Vojvodic, Aleksandra and Yang, Jenny Y. and Balsara, Nitash P. and Persson, Kristin A. and Siegel, Donald J. and Haynes, Christy L. and Mauzeroll, Janine and Shen, Mei and Venton, B. Jill and Balke, Nina and {Rodr{\'i}guez-L{\'o}pez}, Joaqu{\'i}n and Rolison, Debra R. and {Shahbazian-Yassar}, Reza and Srinivasan, Venkat and Chaudhuri, Santanu and Couet, Adrien and {Hattrick-Simpers}, Jason},
  year = 2022,
  month = jan,
  journal = {ACS Energy Lett.},
  volume = {7},
  number = {1},
  pages = {368--374},
  issn = {2380-8195, 2380-8195},
  doi = {10.1021/acsenergylett.1c02608},
  urldate = {2025-05-27},
  copyright = {https://doi.org/10.15223/policy-001},
  langid = {english}
}

@article{Chen2025,
  title = {Prediction of Coating Degradation Based on ``{{Environmental Factors}}--{{Physical Property}}--{{Corrosion Failure}}'' Two-Stage Machine Learning},
  author = {Chen, Weiting and Ma, Lingwei and Li, Yiran and Wu, Dequan and Zhou, Kun and Wang, Jinke and Chen, Zhibin and Guo, Xin and Li, Zongbao and Chowwanonthapunya, Thee and Li, Xiaogang and Zhang, Dawei},
  year = 2025,
  month = jun,
  journal = {npj Mater Degrad},
  volume = {9},
  number = {1},
  pages = {67},
  issn = {2397-2106},
  doi = {10.1038/s41529-025-00614-6},
  urldate = {2025-09-25},
  langid = {english}
}

@article{Christensen2021,
  title = {Automation Isn't Automatic},
  author = {Christensen, Melodie and Yunker, Lars P. E. and Shiri, Parisa and Zepel, Tara and Prieto, Paloma L. and Grunert, Shad and Bork, Finn and Hein, Jason E.},
  year = 2021,
  journal = {Chem. Sci.},
  volume = {12},
  number = {47},
  pages = {15473--15490},
  issn = {2041-6520, 2041-6539},
  doi = {10.1039/D1SC04588A},
  urldate = {2025-10-09},
  langid = {english}
}

@article{Darvish2025,
  title = {{{ORGANA}}: {{A}} Robotic Assistant for Automated Chemistry Experimentation and Characterization},
  shorttitle = {{{ORGANA}}},
  author = {Darvish, Kourosh and Skreta, Marta and Zhao, Yuchi and Yoshikawa, Naruki and Som, Sagnik and Bogdanovic, Miroslav and Cao, Yang and Hao, Han and Xu, Haoping and {Aspuru-Guzik}, Al{\'a}n and Garg, Animesh and Shkurti, Florian},
  year = 2025,
  month = feb,
  journal = {Matter},
  volume = {8},
  number = {2},
  pages = {101897},
  issn = {25902385},
  doi = {10.1016/j.matt.2024.10.015},
  urldate = {2025-04-15},
  langid = {english}
}

@article{Dastgerdi2019,
  title = {Electrochemical Methods for the Determination of {{Pedeferri}}'s Diagram of Stainless Steel in Chloride Containing Environment},
  author = {Dastgerdi, Arash Azimi and Brenna, Andrea and Ormellese, Marco and Pedeferri, MariaPia and Bolzoni, Fabio},
  year = 2019,
  month = jan,
  journal = {Materials \& Corrosion},
  volume = {70},
  number = {1},
  pages = {9--18},
  issn = {0947-5117, 1521-4176},
  doi = {10.1002/maco.201810386},
  urldate = {2025-04-15},
  langid = {english}
}

@phdthesis{Dastgerdi2019a,
  title = {Pedeferri's Diagram of Stainless Steel Type {{AISI 304L}}},
  author = {Dastgerdi, Arash Azimi},
  year = 2019,
  school = {POLITECNICO DI MILANO}
}

@article{DeCost2022,
  title = {Towards {{Automated Design}} of {{Corrosion Resistant Alloy Coatings}} with an {{Autonomous Scanning Droplet Cell}}},
  author = {DeCost, Brian and Joress, Howie and Sarker, Suchismita and Mehta, Apurva and {Hattrick-Simpers}, Jason},
  year = 2022,
  month = aug,
  journal = {JOM},
  volume = {74},
  number = {8},
  pages = {2941--2950},
  issn = {1047-4838, 1543-1851},
  doi = {10.1007/s11837-022-05367-0},
  urldate = {2023-06-16},
  langid = {english}
}

@article{Delahay1950,
  title = {Potential-{{pH}} Diagrams},
  author = {Delahay, Paul and Pourbaix, Marcel and Van Rysselberghe, Pierre},
  year = 1950,
  month = dec,
  journal = {J. Chem. Educ.},
  volume = {27},
  number = {12},
  pages = {683},
  issn = {0021-9584, 1938-1328},
  doi = {10.1021/ed027p683},
  urldate = {2025-10-15},
  langid = {english}
}

@article{Fisker-Bodker2025,
  title = {{{AMPERE-2}}: An Open-Hardware, Robotic Platform for Automated Electrodeposition and Electrochemical Validation},
  shorttitle = {{{AMPERE-2}}},
  author = {{Fisker-B{\o}dker}, Nis and Persaud, Daniel and Bai, Yang and Kozdras, Mark and Vegge, Tejs and {Hattrick-Simpers}, Jason and Chang, Jin Hyun},
  year = 2025,
  journal = {Digital Discovery},
  volume = {4},
  number = {9},
  pages = {2491--2501},
  issn = {2635-098X},
  doi = {10.1039/D5DD00180C},
  urldate = {2025-09-24},
  langid = {english}
}

@misc{Flint1980,
  title = {Resistance of Stainless Steel to Corrosion in Naturally Occurring Waters},
  author = {Flint, G.N.},
  year = 1980,
  publisher = {Nickel Institute}
}

@article{Gerroll2023,
  title = {Legion: {{An Instrument}} for {{High-Throughput Electrochemistry}}},
  shorttitle = {Legion},
  author = {Gerroll, Benjamin H. R. and Kulesa, Krista M. and Ault, Charles A. and Baker, Lane A.},
  year = 2023,
  month = oct,
  journal = {ACS Meas. Sci. Au},
  volume = {3},
  number = {5},
  pages = {371--379},
  issn = {2694-250X, 2694-250X},
  doi = {10.1021/acsmeasuresciau.3c00022},
  urldate = {2023-12-20},
  langid = {english}
}

@inbook{Groysman2010,
  title = {Corrosion {{Monitoring}}},
  booktitle = {Corrosion for {{Everybody}}},
  author = {Groysman, A.},
  year = 2010,
  pages = {189--230},
  publisher = {Springer Netherlands},
  address = {Dordrecht},
  doi = {10.1007/978-90-481-3477-9_5},
  urldate = {2025-05-27},
  collaborator = {Groysman, Alec},
  isbn = {978-90-481-3494-6 978-90-481-3477-9},
  langid = {english}
}

@article{Hariharan2024,
  title = {Microstructure Engineering for Corrosion Resistance in Structural Alloy Design},
  author = {Hariharan, Karthikeyan and Virtanen, Sannakaisa},
  year = 2024,
  month = nov,
  journal = {npj Mater Degrad},
  volume = {8},
  number = {1},
  pages = {115},
  issn = {2397-2106},
  doi = {10.1038/s41529-024-00533-y},
  urldate = {2025-09-25},
  langid = {english}
}

@article{Hashimoto2026,
  title = {Effect of {{Phosphate Addition}} to {{Electrolytes}} on {{Corrosion Behavior}} of {{Stainless Steels}} in {{Seawater Electrolysis}}},
  author = {Hashimoto, Tomoya and Kadowaki, Mariko and Murase, Yoshiharu and Shimabukuro, Masaya and Takanabe, Kazuhiro and Kawashita, Masakazu and Katayama, Hideki and Tsutsumi, Yusuke},
  year = 2026,
  month = apr,
  journal = {ACS Omega},
  volume = {11},
  number = {13},
  pages = {20500--20508},
  issn = {2470-1343, 2470-1343},
  doi = {10.1021/acsomega.5c11989},
  urldate = {2026-05-07},
  copyright = {https://creativecommons.org/licenses/by/4.0/},
  langid = {english}
}

@article{HusseinKhalaf2024,
  title = {Emerging {{AI}} Technologies for Corrosion Monitoring in Oil and Gas Industry: {{A}} Comprehensive Review},
  shorttitle = {Emerging {{AI}} Technologies for Corrosion Monitoring in Oil and Gas Industry},
  author = {Hussein Khalaf, Ali and Xiao, Ying and Xu, Ning and Wu, Bohong and Li, Huan and Lin, Bing and Nie, Zhen and Tang, Junlei},
  year = 2024,
  month = jan,
  journal = {Engineering Failure Analysis},
  volume = {155},
  pages = {107735},
  issn = {13506307},
  doi = {10.1016/j.engfailanal.2023.107735},
  urldate = {2025-05-27},
  langid = {english}
}

@article{Iannuzzi2022,
  title = {The Carbon Footprint of Steel Corrosion},
  author = {Iannuzzi, M. and Frankel, G. S.},
  year = 2022,
  month = dec,
  journal = {npj Mater Degrad},
  volume = {6},
  number = {1},
  pages = {101},
  issn = {2397-2106},
  doi = {10.1038/s41529-022-00318-1},
  urldate = {2025-09-25},
  langid = {english}
}

@article{Jenewein2024,
  title = {Automated Monitoring of Electrocatalyst Corrosion as a Function of Electrochemical History and Electrolyte Formulation},
  author = {Jenewein, Ken J. and Kan, Kevin and Guevarra, Dan and Jones, Ryan J. R. and Lai, Yungchieh and Suram, Santosh K. and Haber, Joel A. and Cherevko, Serhiy and Gregoire, John M.},
  year = 2024,
  journal = {Chem. Commun.},
  volume = {60},
  number = {71},
  pages = {9554--9557},
  issn = {1359-7345, 1364-548X},
  doi = {10.1039/D4CC02906B},
  urldate = {2025-09-24},
  langid = {english}
}

@article{Joress2022,
  title = {Development of an Automated Millifluidic Platform and Data-Analysis Pipeline for Rapid Electrochemical Corrosion Measurements: {{A pH}} Study on {{Zn-Ni}}},
  shorttitle = {Development of an Automated Millifluidic Platform and Data-Analysis Pipeline for Rapid Electrochemical Corrosion Measurements},
  author = {Joress, Howie and DeCost, Brian and Hassan, Najlaa and Braun, Trevor M. and Gorham, Justin M. and {Hattrick-Simpers}, Jason},
  year = 2022,
  month = oct,
  journal = {Electrochimica Acta},
  volume = {428},
  pages = {140866},
  issn = {00134686},
  doi = {10.1016/j.electacta.2022.140866},
  urldate = {2023-12-03},
  langid = {english}
}

@article{Koch2016,
  title = {International Measures of Prevention, Application, and Economics of Corrosion Technologies Study},
  author = {Koch, Gerhardus and Varney, Jeff and Thompson, Neil and Moghissi, Oliver and Gould, Melissa and Payer, Joe},
  year = 2016,
  journal = {NACE int},
  volume = {216},
  number = {3}
}

@incollection{Koch2017,
  title = {Cost of Corrosion},
  booktitle = {Trends in {{Oil}} and {{Gas Corrosion Research}} and {{Technologies}}},
  author = {Koch, Gerhardus},
  year = 2017,
  pages = {3--30},
  publisher = {Elsevier},
  doi = {10.1016/B978-0-08-101105-8.00001-2},
  urldate = {2023-11-29},
  isbn = {978-0-08-101105-8},
  langid = {english}
}

@article{Kolotyrkin1963,
  title = {Pitting {{Corrosion}} of {{Metals}}},
  author = {Kolotyrkin, {\relax Ja}. M.},
  year = 1963,
  month = aug,
  journal = {Corrosion},
  volume = {19},
  number = {8},
  pages = {261t-268t},
  issn = {1938-159X, 0010-9312},
  doi = {10.5006/0010-9312-19.8.261},
  urldate = {2025-11-11},
  langid = {english}
}

@article{Kulesa2024,
  title = {Interfacing {{High-Throughput Electrosynthesis}} and {{Mass Spectrometric Analysis}} of {{Azines}}},
  author = {Kulesa, Krista M. and Hirtzel, Erin A. and Nguyen, Vinh T. and Freitas, Dallas P. and Edwards, Madison E. and Yan, Xin and Baker, Lane A.},
  year = 2024,
  month = may,
  journal = {Anal. Chem.},
  volume = {96},
  number = {21},
  pages = {8249--8253},
  issn = {0003-2700, 1520-6882},
  doi = {10.1021/acs.analchem.4c01110},
  urldate = {2025-10-14},
  copyright = {https://creativecommons.org/licenses/by/4.0/},
  langid = {english}
}

@article{Lakatos-Varsanyi1998,
  title = {The Influence of Phosphate on Repassivation of 304 Stainless Steel in Neutral Chloride Solution},
  author = {{Lakatos-Vars{\'a}nyi}, M. and Falkenberg, F. and Olefjord, I.},
  year = 1998,
  month = jan,
  journal = {Electrochimica Acta},
  volume = {43},
  number = {1-2},
  pages = {187--197},
  issn = {00134686},
  doi = {10.1016/S0013-4686(97)00224-7},
  urldate = {2026-05-07},
  copyright = {https://www.elsevier.com/tdm/userlicense/1.0/},
  langid = {english}
}

@article{Mongay1974,
  title = {A {{Britton-Robinson Buffer}} of {{Known Ionic Strength}}},
  author = {{Carlos Mongay} and {Victor Cerda}},
  year = 1974,
  journal = {Ann. Chim},
  volume = {64},
  number = {5},
  pages = {5}
}

@article{Nnoka2024,
  title = {Effects of Different Microstructural Parameters on the Corrosion and Cracking Resistance of Pipeline Steels: {{A}} Review},
  shorttitle = {Effects of Different Microstructural Parameters on the Corrosion and Cracking Resistance of Pipeline Steels},
  author = {Nnoka, Meekness and Alaso Jack, Tonye and Szpunar, Jerzy},
  year = 2024,
  month = may,
  journal = {Engineering Failure Analysis},
  volume = {159},
  pages = {108065},
  issn = {13506307},
  doi = {10.1016/j.engfailanal.2024.108065},
  urldate = {2025-09-25},
  langid = {english}
}

@article{Nwabara2025,
  title = {High Throughput Computational and Experimental Methods for Accelerated Electrochemical Materials Discovery},
  author = {Nwabara, Uzoma and Yang, Kunran and Talekar, Akshay and Bernales, Varinia and Gonz{\'a}lez, Jorge and Miller, Stuart and Wu, Jinfeng},
  year = 2025,
  journal = {J. Mater. Chem. A},
  volume = {13},
  number = {32},
  pages = {26041--26066},
  issn = {2050-7488, 2050-7496},
  doi = {10.1039/D5TA00331H},
  urldate = {2025-10-14},
  langid = {english}
}

@article{Oh2023,
  title = {The {{Electrolab}}: {{An}} Open-Source, Modular Platform for Automated Characterization of Redox-Active Electrolytes},
  shorttitle = {The {{Electrolab}}},
  author = {Oh, Inkyu and Pence, Michael A. and Lukhanin, Nikita G. and Rodr{\'i}guez, Oliver and Schroeder, Charles M. and {Rodr{\'i}guez-L{\'o}pez}, Joaqu{\'i}n},
  year = 2023,
  month = nov,
  journal = {Device},
  volume = {1},
  number = {5},
  pages = {100103},
  issn = {26669986},
  doi = {10.1016/j.device.2023.100103},
  urldate = {2025-12-03},
  langid = {english}
}

@article{Pedeferri1996,
  title = {Cathodic Protection and Cathodic Prevention},
  author = {Pedeferri, Pietro},
  year = 1996,
  month = jul,
  journal = {Construction and Building Materials},
  volume = {10},
  number = {5},
  pages = {391--402},
  issn = {09500618},
  doi = {10.1016/0950-0618(95)00017-8},
  urldate = {2025-10-15},
  langid = {english}
}

@article{Pence2025,
  title = {The Emergence of Automation in Electrochemistry},
  author = {Pence, Michael A. and Hazen, Gavin and {Rodr{\'i}guez-L{\'o}pez}, Joaqu{\'i}n},
  year = 2025,
  month = jun,
  journal = {Current Opinion in Electrochemistry},
  volume = {51},
  pages = {101679},
  issn = {24519103},
  doi = {10.1016/j.coelec.2025.101679},
  urldate = {2025-05-27},
  langid = {english}
}

@article{Pimenova2008,
  title = {Electrochemical {{Corrosion Investigation}} of 49-{{Cell Combinatorial Library}} of {{Titanium-Based Alloys Fabricated}} by {{DMD}}},
  author = {Pimenova, Natalia and Starr, Thomas L.},
  year = 2008,
  journal = {J. Electrochem. Soc.},
  volume = {155},
  number = {6},
  pages = {C303},
  issn = {00134651},
  doi = {10.1149/1.2899021},
  urldate = {2025-09-25},
  langid = {english}
}

@article{Pistorius1992,
  title = {Metastable Pitting Corrosion of Stainless Steel and the Transition to Stability},
  author = {Pistorius, P. C. and Burstein, G. T.},
  year = 1992,
  month = dec,
  journal = {Phil. Trans. R. Soc. Lond. A},
  volume = {341},
  number = {1662},
  pages = {531--559},
  issn = {0962-8428, 2054-0299},
  doi = {10.1098/rsta.1992.0114},
  urldate = {2025-11-07},
  copyright = {https://royalsociety.org/journals/ethics-policies/data-sharing-mining/},
  langid = {english}
}

@article{Pistorius1994,
  title = {Aspects of the Effects of Electrolyte Composition on the Occurrence of Metastable Pitting on Stainless Steel},
  author = {Pistorius, P.C. and Burstein, G.T.},
  year = 1994,
  month = mar,
  journal = {Corrosion Science},
  volume = {36},
  number = {3},
  pages = {525--538},
  issn = {0010938X},
  doi = {10.1016/0010-938X(94)90041-8},
  urldate = {2026-05-07},
  copyright = {https://www.elsevier.com/tdm/userlicense/1.0/},
  langid = {english}
}

@article{Quinn2024,
  title = {{{PANDA}}: A Self-Driving Lab for Studying Electrodeposited Polymer Films},
  shorttitle = {{{PANDA}}},
  author = {Quinn, Harley and Robben, Gregory A. and Zheng, Zhaoyi and Gardner, Alan L. and Werner, J{\"o}rg G. and Brown, Keith A.},
  year = 2024,
  journal = {Mater. Horiz.},
  volume = {11},
  number = {21},
  pages = {5331--5340},
  issn = {2051-6347, 2051-6355},
  doi = {10.1039/D4MH00797B},
  urldate = {2025-04-10},
  langid = {english}
}

@article{Rial-Rodriguez2024,
  title = {An {{Automated Electrochemical Flow Platform}} to {{Accelerate Library Synthesis}} and {{Reaction Optimization}}},
  author = {Rial-Rodr{\'i}guez, Eduardo and Williams, Jason D. and Cantillo, David and Fuch{\ss}, Thomas and Sommer, Alena and Eggenweiler, Hans-Michael and Kappe, C. Oliver and Laudadio, Gabriele},
  year = 2024,
  month = dec,
  journal = {Angew Chem Int Ed},
  volume = {63},
  number = {51},
  pages = {e202412045},
  issn = {1433-7851, 1521-3773},
  doi = {10.1002/anie.202412045},
  urldate = {2025-12-03},
  langid = {english}
}

@article{Salvarezza1987,
  title = {Stochastic and Deterministic Behaviours of 316 Stainless Steel Pitting Corrosion in Phosphate-Borate Buffer Containing Sodium Chloride},
  author = {Salvarezza, R.C. and De Cristofaro, N. and Pallotta, C. and Arvia, A.J.},
  year = 1987,
  month = jul,
  journal = {Electrochimica Acta},
  volume = {32},
  number = {7},
  pages = {1049--1055},
  issn = {00134686},
  doi = {10.1016/0013-4686(87)90032-6},
  urldate = {2026-05-07},
  copyright = {https://www.elsevier.com/tdm/userlicense/1.0/},
  langid = {english}
}

@article{Sheng2024,
  title = {Autonomous Closed-Loop Mechanistic Investigation of Molecular Electrochemistry via Automation},
  author = {Sheng, Hongyuan and Sun, Jingwen and Rodr{\'i}guez, Oliver and Hoar, Benjamin B. and Zhang, Weitong and Xiang, Danlei and Tang, Tianhua and Hazra, Avijit and Min, Daniel S. and Doyle, Abigail G. and Sigman, Matthew S. and Costentin, Cyrille and Gu, Quanquan and {Rodr{\'i}guez-L{\'o}pez}, Joaqu{\'i}n and Liu, Chong},
  year = 2024,
  month = mar,
  journal = {Nat Commun},
  volume = {15},
  number = {1},
  pages = {2781},
  issn = {2041-1723},
  doi = {10.1038/s41467-024-47210-x},
  urldate = {2025-04-15},
  langid = {english}
}

@article{Shi2016,
  title = {A {{Smart High-Throughput Experiment Platform}} for {{Materials Corrosion Study}}},
  author = {Shi, Peng and Li, Bin and Huo, Jindong and Wen, Lei},
  year = 2016,
  journal = {Scientific Programming},
  volume = {2016},
  pages = {1--9},
  issn = {1058-9244, 1875-919X},
  doi = {10.1155/2016/6876241},
  urldate = {2025-09-24},
  copyright = {http://creativecommons.org/licenses/by/4.0/},
  langid = {english}
}

@article{Stier2024,
  title = {Materials {{Acceleration Platforms}} ({{MAPs}}): {{Accelerating Materials Research}} and {{Development}} to {{Meet Urgent Societal Challenges}}},
  shorttitle = {Materials {{Acceleration Platforms}} ({{MAPs}})},
  author = {Stier, Simon P. and Kreisbeck, Christoph and Ihssen, Holger and Popp, Matthias Albert and Hauch, Jens and Malek, Kourosh and Reynaud, Marine and Goumans, T.P.M. and Carlsson, Johan and Todorov, Ilian and Gold, Lukas and R{\"a}der, Andreas and Wenzel, Wolfgang and Bandesha, Shahbaz Tareq and Jacques, Philippe and Garcia-Moreno, Francisco and Arcelus, Oier and Friederich, Pascal and Clark, Simon and Maglione, Mario and Laukkanen, Anssi and Castelli, Ivano Eligio and Carrasco, Javier and Cabanas, Montserrat Casas and Stein, Helge S{\"o}ren and Ozcan, Ozlem and Elbert, David and Reuter, Karsten and Scheurer, Christoph and Demura, Masahiko and Han, Sang Soo and Vegge, Tejs and Nakamae, Sawako and Fabrizio, Monica and Kozdras, Mark},
  year = 2024,
  month = nov,
  journal = {Advanced Materials},
  volume = {36},
  number = {45},
  pages = {2407791},
  issn = {0935-9648, 1521-4095},
  doi = {10.1002/adma.202407791},
  urldate = {2025-09-26},
  langid = {english}
}

@article{Taylor2011,
  title = {The Investigation of Corrosion Phenomena with High Throughput Methods: A Review},
  shorttitle = {The Investigation of Corrosion Phenomena with High Throughput Methods},
  author = {Taylor, S. Ray},
  year = 2011,
  month = oct,
  journal = {Corrosion Reviews},
  volume = {29},
  number = {3-4},
  pages = {135--151},
  issn = {2191-0316, 0334-6005},
  doi = {10.1515/CORRREV.2011.024},
  urldate = {2025-09-25},
  langid = {english}
}

@article{Wang2019,
  title = {Effect of {{pH}} on the {{Electrochemical Behaviour}} and {{Passive Film Composition}} of {{316L Stainless Steel}}},
  author = {Wang, Zhu and Zhou, Zi-Qiang and Zhang, Lei and Hu, Jia-Yuan and Zhang, Zi-Ru and Lu, Min-Xu},
  year = 2019,
  month = may,
  journal = {Acta Metall. Sin. (Engl. Lett.)},
  volume = {32},
  number = {5},
  pages = {585--598},
  issn = {1006-7191, 2194-1289},
  doi = {10.1007/s40195-018-0794-5},
  urldate = {2025-11-11},
  langid = {english}
}

@article{Wetzel2022,
  title = {The Comparison of the Corrosion Behavior of the {{CrCoNi}} Medium Entropy Alloy and {{CrMnFeCoNi}} High Entropy Alloy},
  author = {Wetzel, Annica and Von Der Au, Marcus and Dietrich, Paul M. and Radnik, J{\"o}rg and Ozcan, Ozlem and Witt, Julia},
  year = 2022,
  month = nov,
  journal = {Applied Surface Science},
  volume = {601},
  pages = {154171},
  issn = {01694332},
  doi = {10.1016/j.apsusc.2022.154171},
  urldate = {2023-12-03},
  langid = {english}
}

@article{White2012,
  title = {A New High-Throughput Method for Corrosion Testing},
  author = {White, P.A. and Smith, G.B. and Harvey, T.G. and Corrigan, P.A. and Glenn, M.A. and Lau, D. and Hardin, S.G. and Mardel, J. and Markley, T.A. and Muster, T.H. and Sherman, N. and Garcia, S.J. and Mol, J.M.C. and Hughes, A.E.},
  year = 2012,
  month = may,
  journal = {Corrosion Science},
  volume = {58},
  pages = {327--331},
  issn = {0010938X},
  doi = {10.1016/j.corsci.2012.01.016},
  urldate = {2025-09-25},
  langid = {english}
}

@article{Wu2018,
  title = {The {{Critical Pitting Chloride Concentration}} of {{Various Stainless Steels Measured}} by an {{Electrochemical Method}}},
  author = {Wu, Xuan and Sun, Yangting and Liu, Yuanyuan and Yang, Yuanyuan and Li, Jin and Jiang, Yiming},
  year = 2018,
  journal = {J. Electrochem. Soc.},
  volume = {165},
  number = {14},
  pages = {C939-C949},
  issn = {0013-4651, 1945-7111},
  doi = {10.1149/2.0861814jes},
  urldate = {2026-05-07},
  langid = {english}
}

@article{Xia2022,
  title = {Electrochemical Measurements Used for Assessment of Corrosion and Protection of Metallic Materials in the Field: {{A}} Critical Review},
  shorttitle = {Electrochemical Measurements Used for Assessment of Corrosion and Protection of Metallic Materials in the Field},
  author = {Xia, Da-Hai and Deng, Cheng-Man and Macdonald, Digby and Jamali, Sina and Mills, Douglas and Luo, Jing-Li and Strebl, Michael G. and Amiri, Mehdi and Jin, Weixian and Song, Shizhe and Hu, Wenbin},
  year = 2022,
  month = jun,
  journal = {Journal of Materials Science \& Technology},
  volume = {112},
  pages = {151--183},
  issn = {10050302},
  doi = {10.1016/j.jmst.2021.11.004},
  urldate = {2025-05-27},
  langid = {english}
}

@article{Zeng2021,
  title = {Effect of Dissolved Oxygen on Electrochemical Corrosion Behavior of 2205 Duplex Stainless Steel in Hot Concentrated Seawater},
  author = {Zeng, Hongtao and Yang, Yong and Zeng, Minhang and Li, Moucheng},
  year = 2021,
  month = mar,
  journal = {Journal of Materials Science \& Technology},
  volume = {66},
  pages = {177--185},
  issn = {10050302},
  doi = {10.1016/j.jmst.2020.06.030},
  urldate = {2025-11-07},
  langid = {english}
}
\newpage

\section{Corrosion Validation Potentiodynamic Polarization}
\label{si:corrosionValidation}
Below are the 32 potentiodynamic polarization (PP) scans and automated analysis to extract the corrosion potential ($E_{\mathrm{corr}}$) and pitting potential ($E_{\mathrm{pit}}$), for each of the ASTM G61-analog experiments performed to validate the MAP-E. The $E_{\mathrm{pit}}$ values extracted from the automated analysis closely match the onset of stable pitting observed in the raw PP scans, all within $\approx$50 mV.

\newpage
\begin{figure}[H]
    \centering
    \textbf{Potentodynamic Polarization scans for batch 1 of replication experiments with the MAP-E.}\par\medskip
    \includegraphics[width=0.80\textwidth]{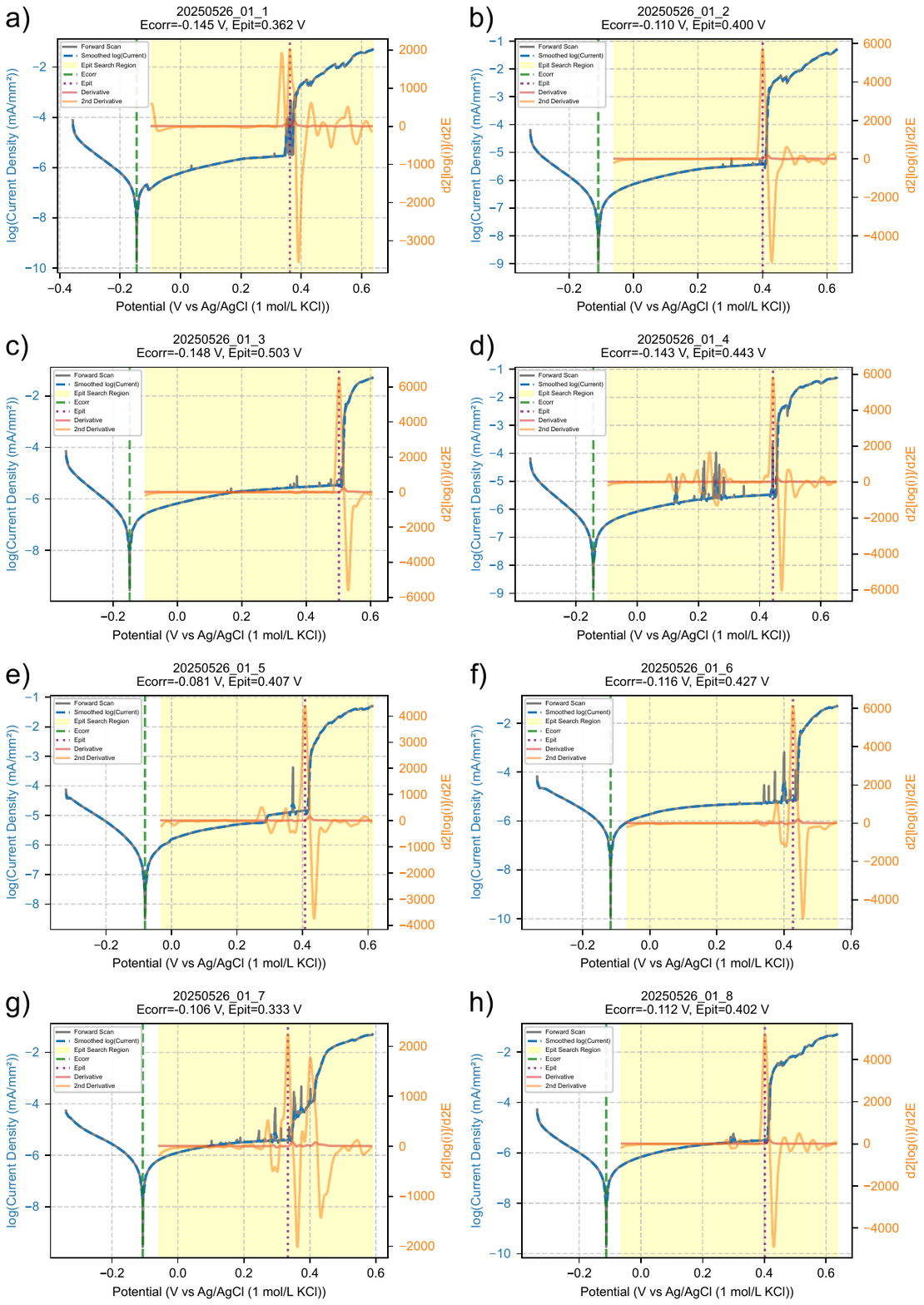}
    \caption{a) cell 1, b) cell 2, c) cell 3, d) cell 4, e) cell 5, f) cell 6, g) cell 7, h) cell 8. Each plot shows the logarithm of the current density vs. potential. The solid grey line is the raw data, blue dashed line is the smoothed data, solid red line is the smoothed data derivative, solid orange line is the smoothed data second derivative, dashed green line is the extracted $E_{\mathrm{corr}}$, and dotted purple line is the extracted $E_{\mathrm{pit}}$.}
    \label{si:corrosionValidation1}
\end{figure}

\newpage
\begin{figure}[H]
    \centering
    \textbf{Potentodynamic Polarization scans for batch 2 of replication experiments with the MAP-E.}\par\medskip
    \includegraphics[width=0.80\textwidth]{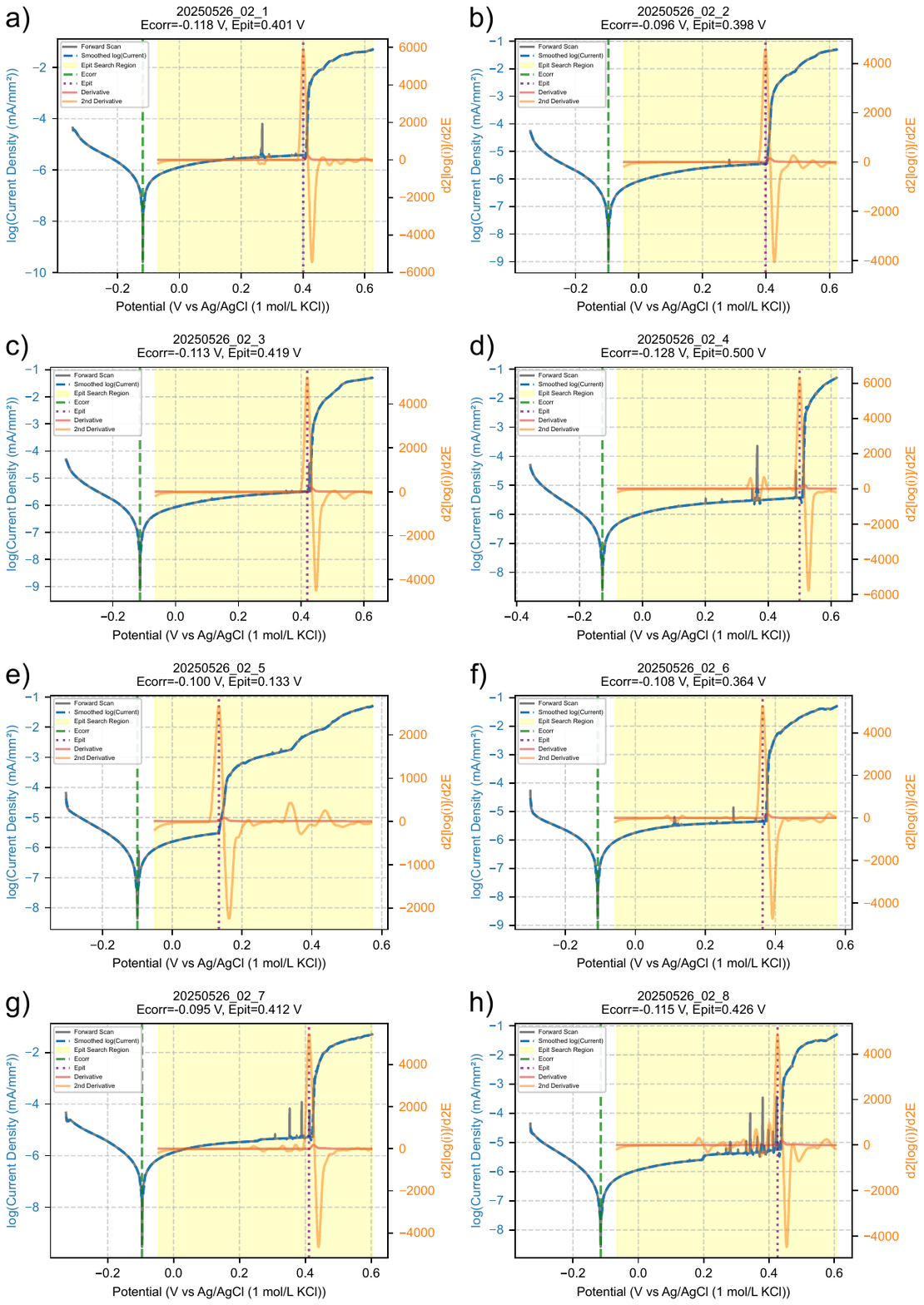}
    \caption{a) cell 1, b) cell 2, c) cell 3, d) cell 4, e) cell 5, f) cell 6, g) cell 7, h) cell 8. Each plot shows the logarithm of the current density vs. potential. The solid grey line is the raw data, blue dashed line is the smoothed data, solid red line is the smoothed data derivative, solid orange line is the smoothed data second derivative, dashed green line is the extracted $E_{\mathrm{corr}}$, and dotted purple line is the extracted $E_{\mathrm{pit}}$.}
    \label{si:corrosionValidation2}
\end{figure}

\newpage
\begin{figure}[H]
    \centering
    \textbf{Potentodynamic Polarization scans for batch 3 of replication experiments with the MAP-E.}\par\medskip
    \includegraphics[width=0.80\textwidth]{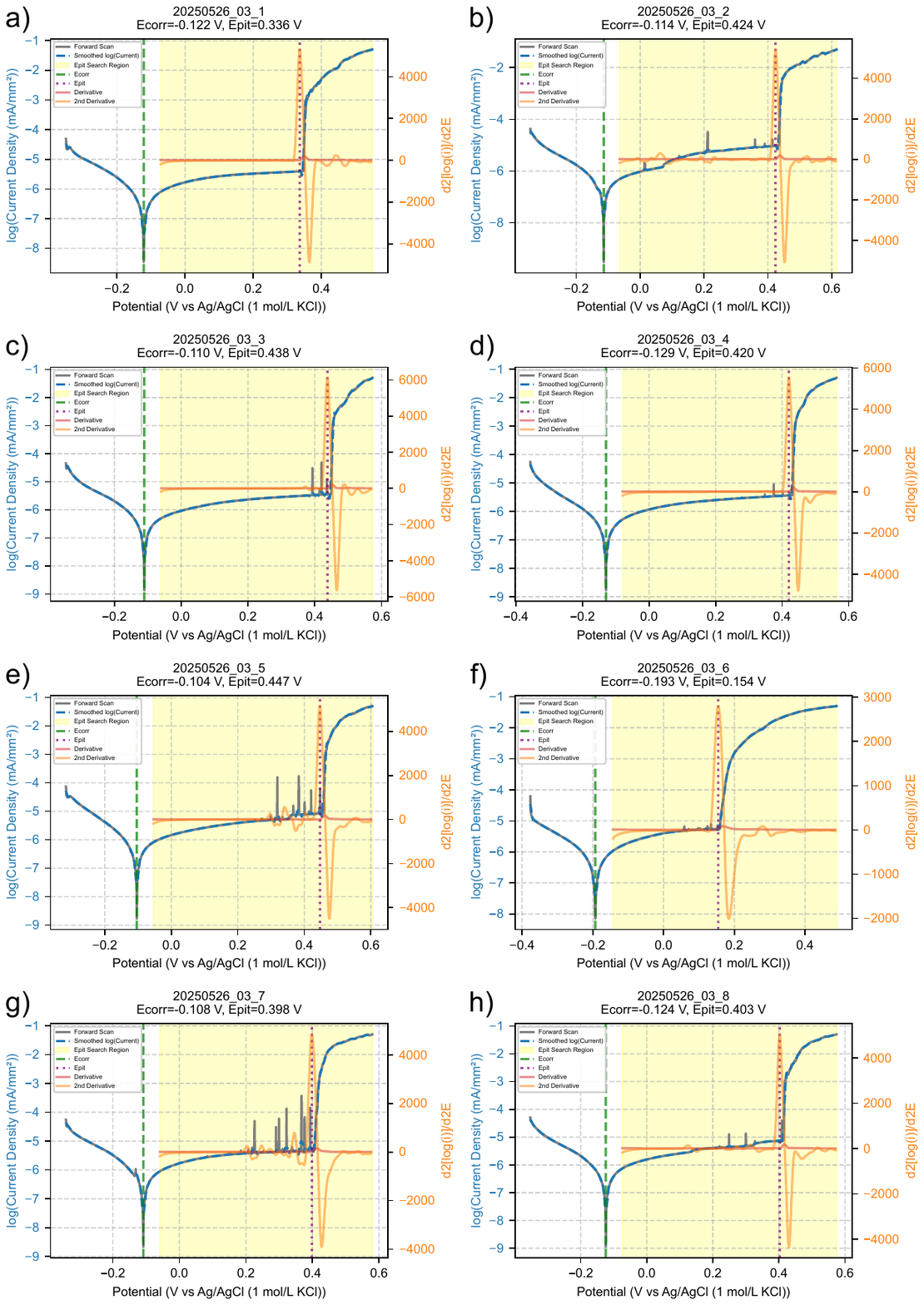}
    \caption{a) cell 1, b) cell 2, c) cell 3, d) cell 4, e) cell 5, f) cell 6, g) cell 7, h) cell 8. Each plot shows the logarithm of the current density vs. potential. The solid grey line is the raw data, blue dashed line is the smoothed data, solid red line is the smoothed data derivative, solid orange line is the smoothed data second derivative, dashed green line is the extracted $E_{\mathrm{corr}}$, and dotted purple line is the extracted $E_{\mathrm{pit}}$.}
    \label{si:corrosionValidation3}
\end{figure}

\newpage
\begin{figure}[H]
    \centering
    \textbf{Potentodynamic Polarization scans for batch 4 of replication experiments with the MAP-E.}\par\medskip
    \includegraphics[width=0.80\textwidth]{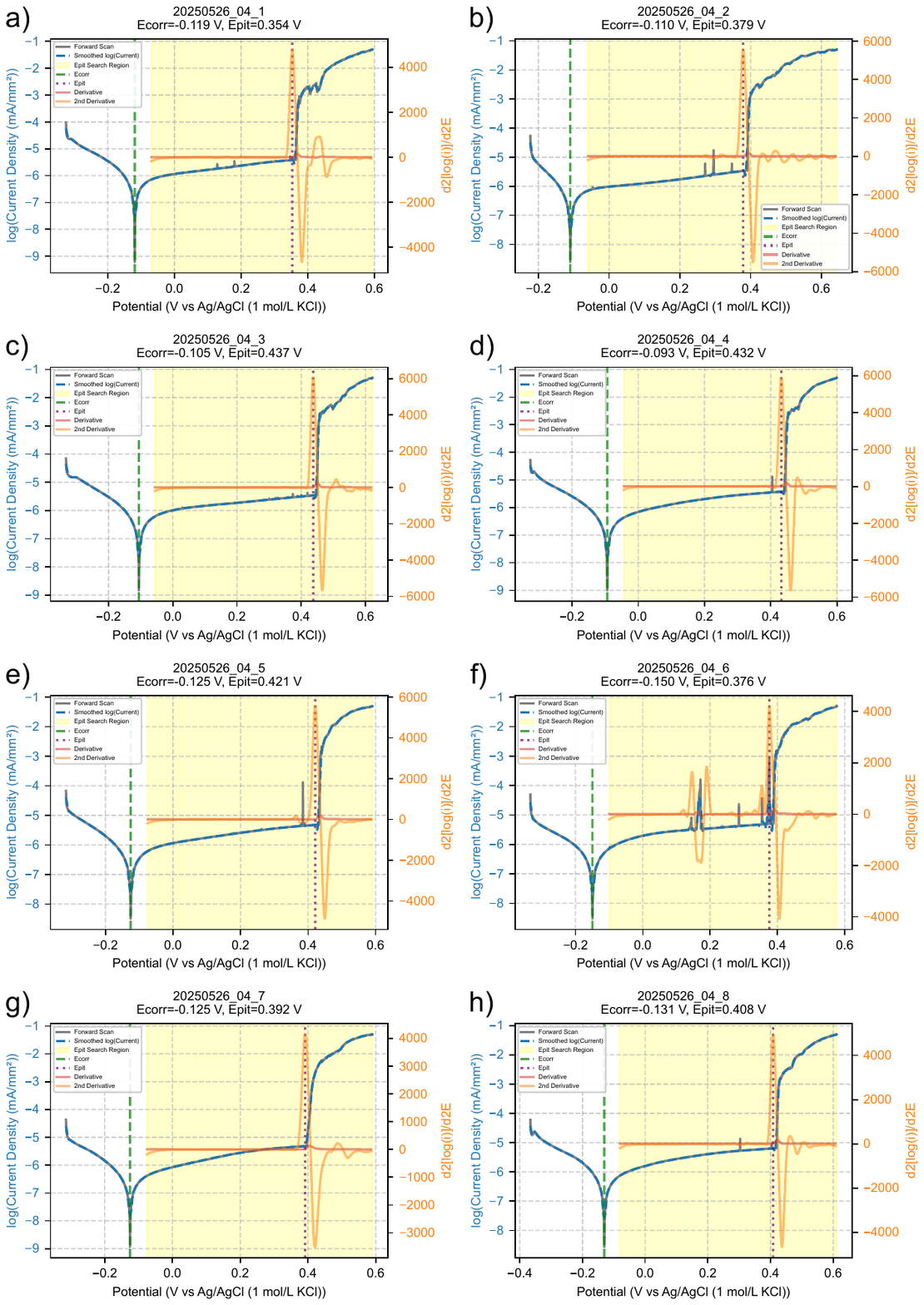}
    \caption{a) cell 1, b) cell 2, c) cell 3, d) cell 4, e) cell 5, f) cell 6, g) cell 7, h) cell 8. Each plot shows the logarithm of the current density vs. potential. The solid grey line is the raw data, blue dashed line is the smoothed data, solid red line is the smoothed data derivative, solid orange line is the smoothed data second derivative, dashed green line is the extracted $E_{\mathrm{corr}}$, and dotted purple line is the extracted $E_{\mathrm{pit}}$.}
    \label{si:corrosionValidation4}
\end{figure}

\newpage
\begin{figure}[H]
    \centering
    \textbf{Potentodynamic Polarization scans for replication experiments with the MAP-E grouped by cell.}\par\medskip
    \includegraphics[width=0.80\textwidth]{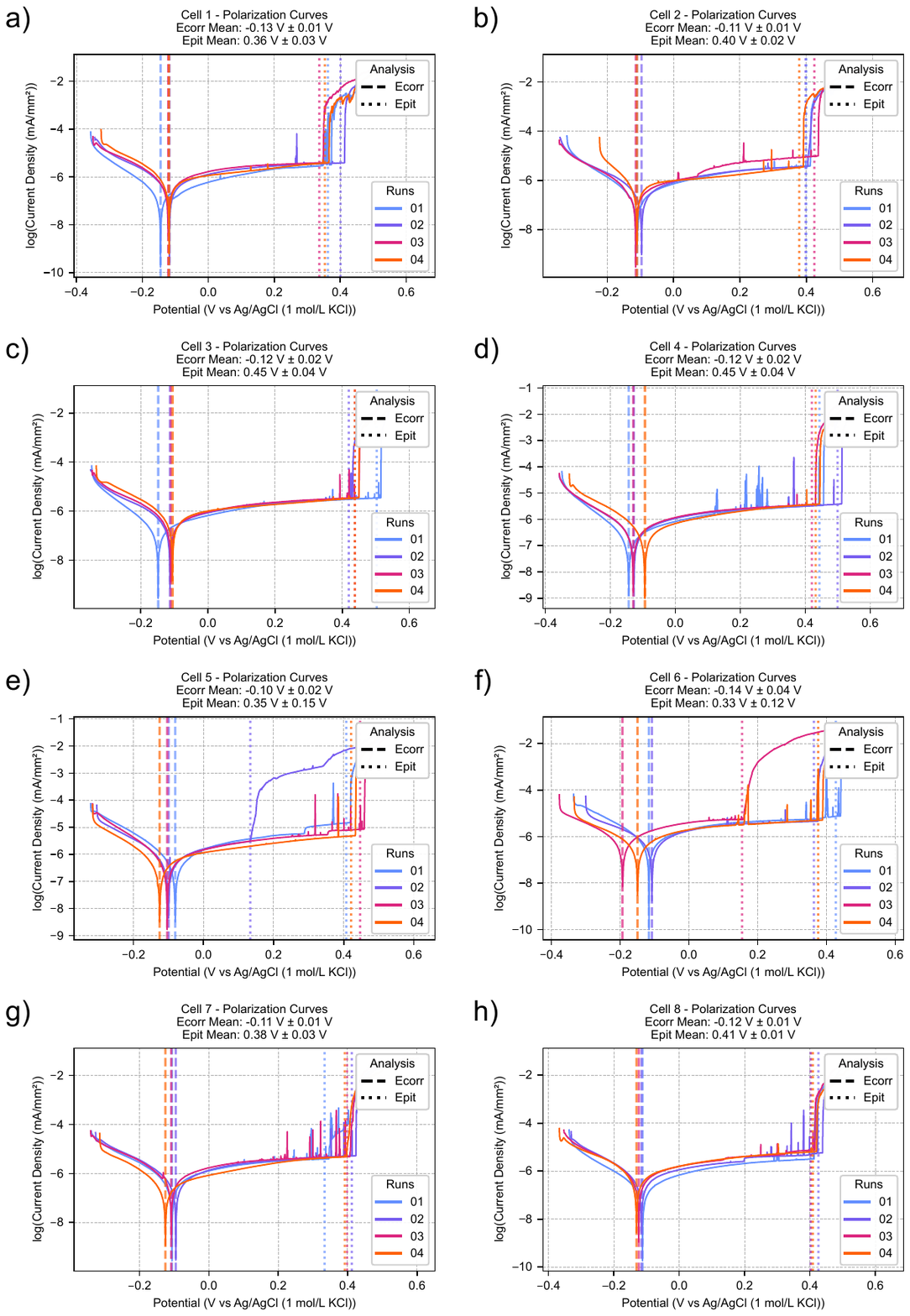}
    \caption{a) cell 1, b) cell 2, c) cell 3, d) cell 4, e) cell 5, f) cell 6, g) cell 7, h) cell 8. Each plot shows the logarithm of the current density vs. potential. The blue solid line is scan 1, purple solid line is scan 2, pink solid line is scan 3, and orange solid line is scan 4. The dashed lines are the E$_{\mathrm{corr}}$ and dotted lines are the E$_{\mathrm{pit}}$ for each scan, with the same color coding.}
    \label{si:corrosionValidationCells}
\end{figure}

\newpage
\section{Autonomous Stability Diagram Potentodynamic Polarization}
\label{si:autonomousStabilityDiagramPPScans}

Below are all potentodynamic polarization (PP) scans and automated analyses used to construct the autonomous stability diagram shown in \hyperref[figure:stabilityDiagramResults]{Figure 3} of the main text, including the 4 initial conditions (8 scans total) and 6 iterations of adaptive experimentation (4 conditions per round, 2 scans per condition), for a total of 56 PP scans.

Similar to the validation experiments in \hyperref[si:corrosionValidation]{section S1}, the automated $E_{\mathrm{pit}}$ extraction closely matched the visually identifiable onset of stable pitting in the majority of cases, with 85\% of extracted values falling within approximately 50 mV of manual interpretation.

In a small number of cases, the automated analysis identified noise fluctuations or secondary current increases instead of the primary pitting transition. Specifically, two scans:

\begin{itemize}
    \item 20250922\_004\_6
    \item 20250922\_004\_7
\end{itemize}

\noindent exhibited noise-related misidentification, while six additional scans:

\begin{itemize}
    \item 20250812\_004
    \item 20250903\_002
    \item 20250916\_004
\end{itemize}

\noindent corresponded to either secondary current rises or scans where stable pitting was not clearly observed.

These behaviors represent known limitations of the current automated extraction algorithm. However, they did not significantly affect the adaptive experimentation workflow because duplicate measurements were performed at each condition and disagreement between replicates was incorporated into the uncertainty estimation used by the GP model. Overall, the extracted values remained sufficient to capture the dominant trends in the stability diagram.

\newpage
\begin{figure}[H]
    \centering
    \textbf{Potentodynamic Polarization scans for iteration 0 of stability diagram building}\par\medskip
    \includegraphics[width=0.80\textwidth]{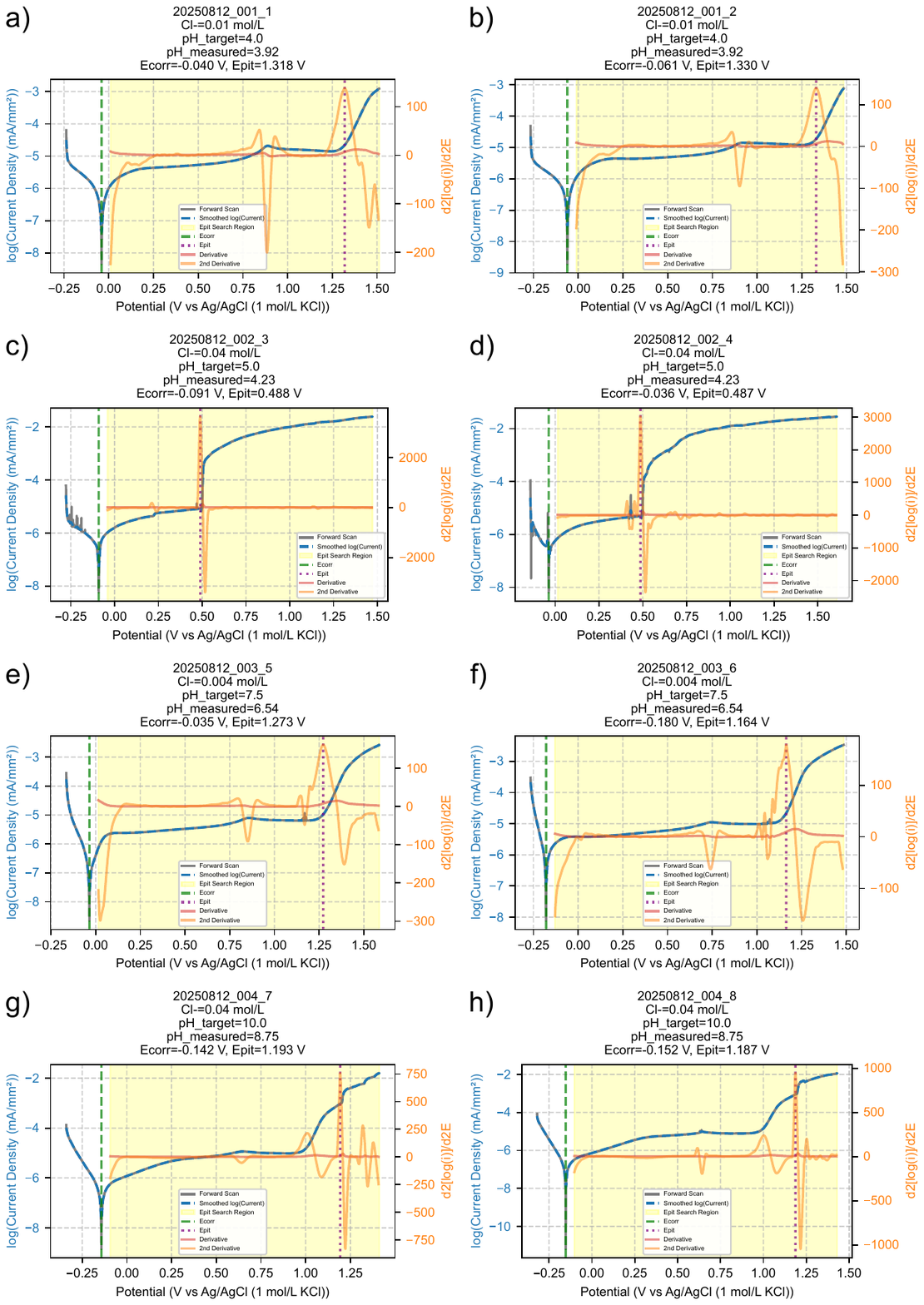}
    \caption{a) cell 1, b) cell 2, c) cell 3, d) cell 4, e) cell 5, f) cell 6, g) cell 7, h) cell 8. Each plot shows the logarithm of the current density vs. potential. The solid grey line is the raw data, blue dashed line is the smoothed data, solid red line is the smoothed data derivative, solid orange line is the smoothed data second derivative, dashed green line is the extracted $E_{\mathrm{corr}}$, and dotted purple line is the extracted $E_{\mathrm{pit}}$.}
    \label{si:autonomousStabilityDiagramScans0}
\end{figure}

\newpage
\begin{figure}[H]
    \centering
    \textbf{Potentodynamic Polarization scans for iteration 1 of stability diagram building}\par\medskip
    \includegraphics[width=0.80\textwidth]{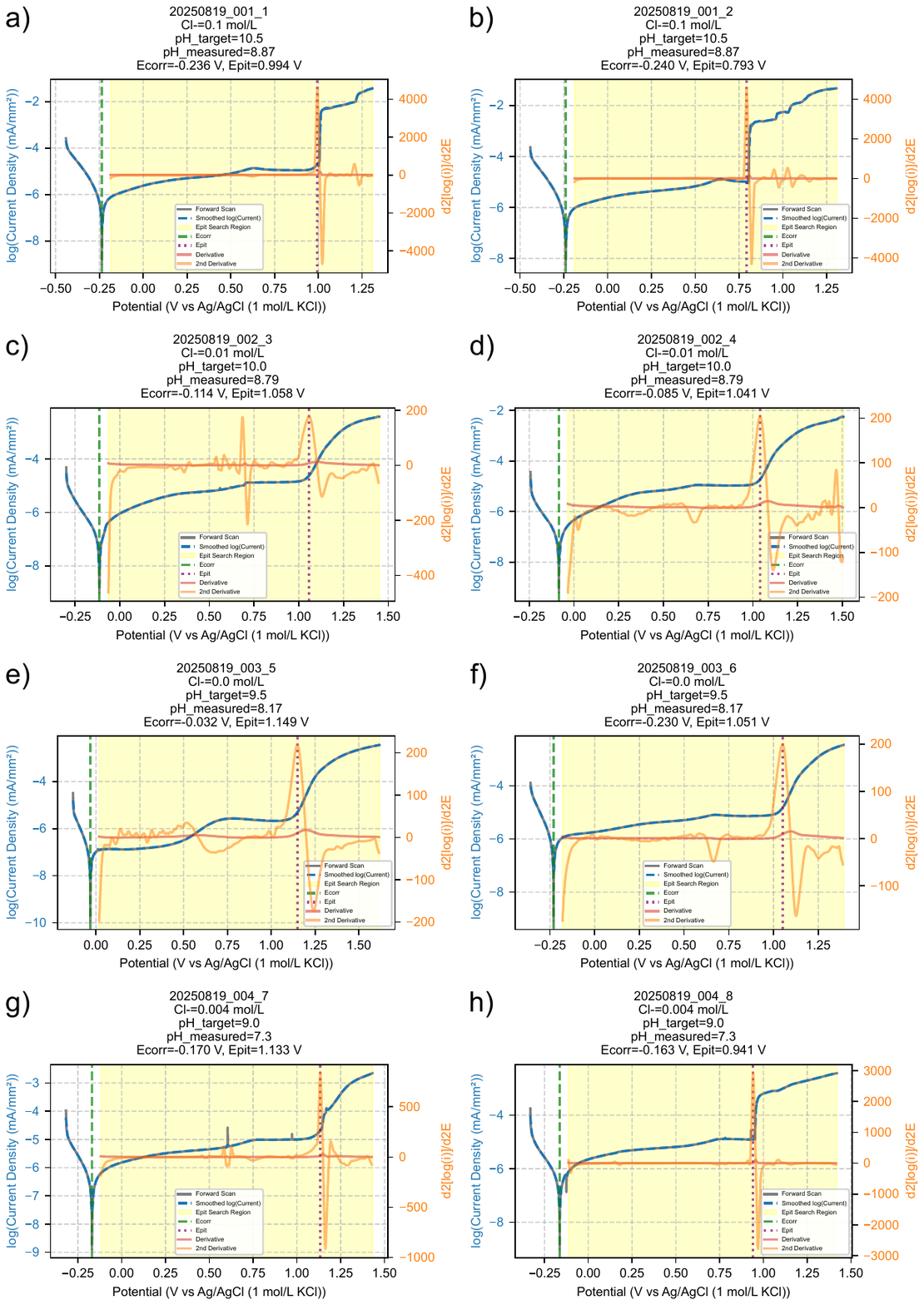}
    \caption{a) cell 1, b) cell 2, c) cell 3, d) cell 4, e) cell 5, f) cell 6, g) cell 7, h) cell 8. Each plot shows the logarithm of the current density vs. potential. The solid grey line is the raw data, blue dashed line is the smoothed data, solid red line is the smoothed data derivative, solid orange line is the smoothed data second derivative, dashed green line is the extracted $E_{\mathrm{corr}}$, and dotted purple line is the extracted $E_{\mathrm{pit}}$.}
    \label{si:autonomousStabilityDiagramScans1}
\end{figure}

\newpage
\begin{figure}[H]
    \centering
    \textbf{Potentodynamic Polarization scans for iteration 2 of stability diagram building}\par\medskip
    \includegraphics[width=0.80\textwidth]{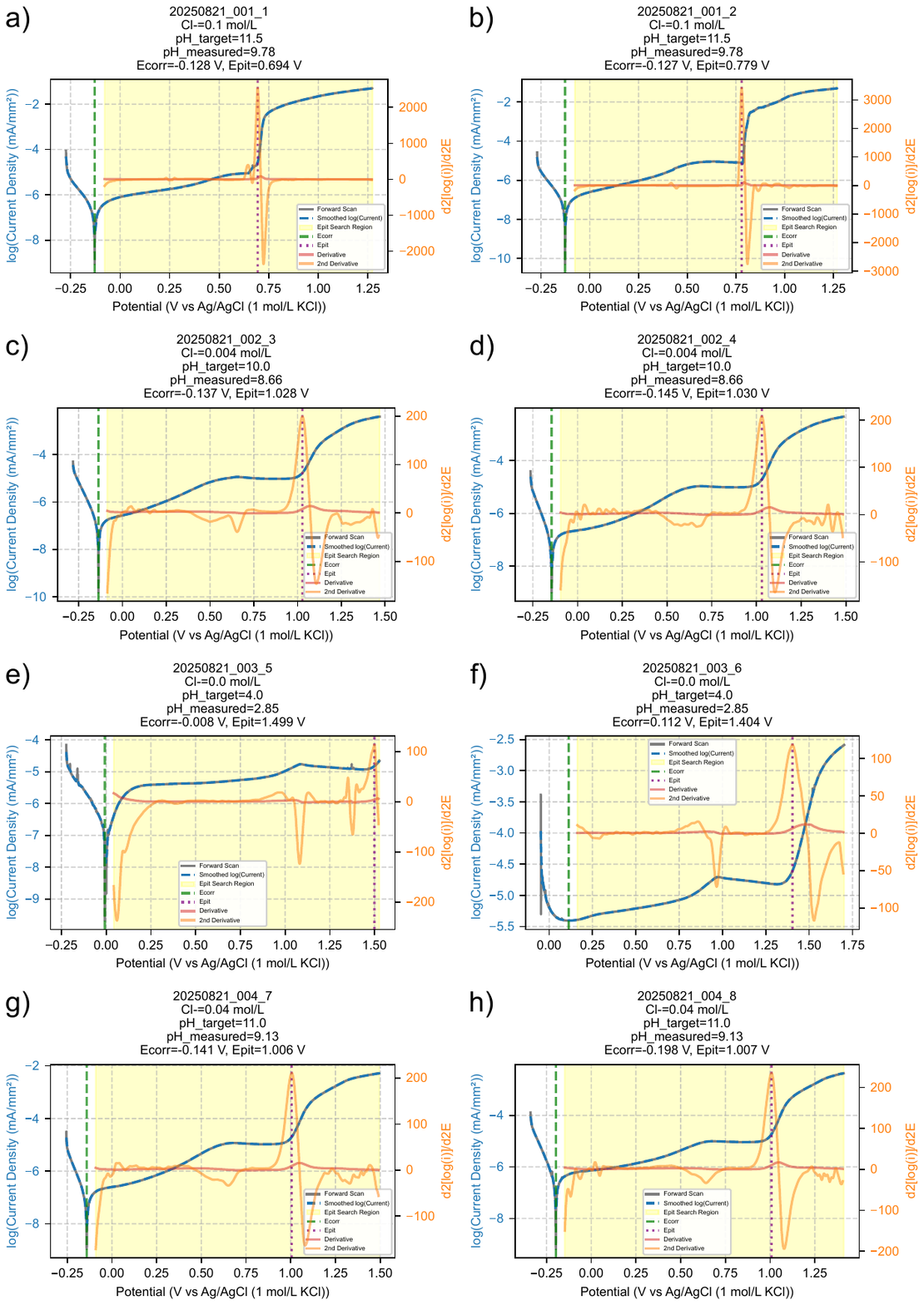}
    \caption{a) cell 1, b) cell 2, c) cell 3, d) cell 4, e) cell 5, f) cell 6, g) cell 7, h) cell 8. Each plot shows the logarithm of the current density vs. potential. The solid grey line is the raw data, blue dashed line is the smoothed data, solid red line is the smoothed data derivative, solid orange line is the smoothed data second derivative, dashed green line is the extracted $E_{\mathrm{corr}}$, and dotted purple line is the extracted $E_{\mathrm{pit}}$.}
    \label{si:autonomousStabilityDiagramScans2}
\end{figure}

\newpage
\begin{figure}[H]
    \centering
    \textbf{Potentodynamic Polarization scans for iteration 3 of stability diagram building}\par\medskip
    \includegraphics[width=0.80\textwidth]{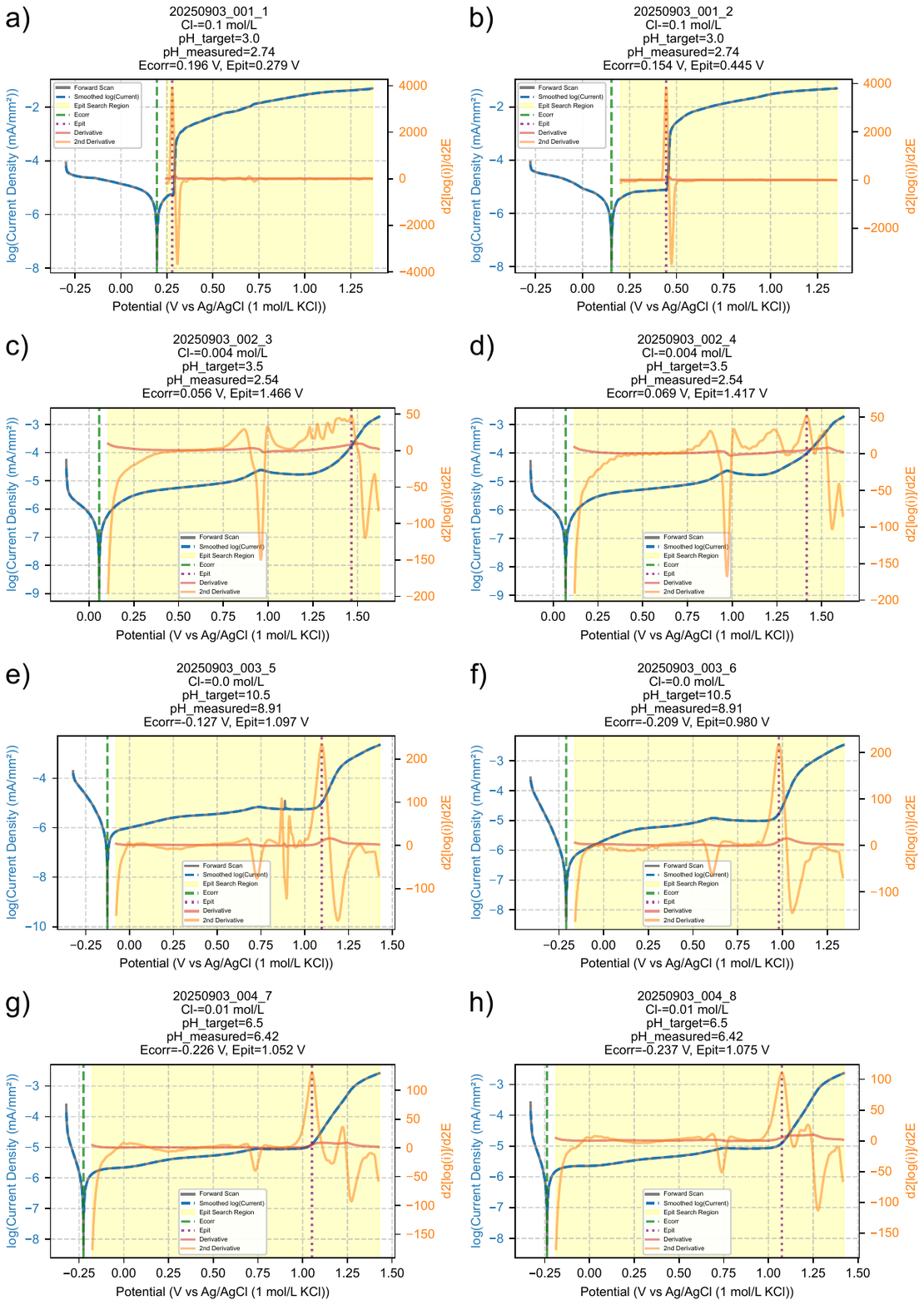}
    \caption{a) cell 1, b) cell 2, c) cell 3, d) cell 4, e) cell 5, f) cell 6, g) cell 7, h) cell 8. Each plot shows the logarithm of the current density vs. potential. The solid grey line is the raw data, blue dashed line is the smoothed data, solid red line is the smoothed data derivative, solid orange line is the smoothed data second derivative, dashed green line is the extracted $E_{\mathrm{corr}}$, and dotted purple line is the extracted $E_{\mathrm{pit}}$.}
    \label{si:autonomousStabilityDiagramScans3}
\end{figure}

\newpage
\begin{figure}[H]
    \centering
    \textbf{Potentodynamic Polarization scans for iteration 4 of stability diagram building}\par\medskip
    \includegraphics[width=0.80\textwidth]{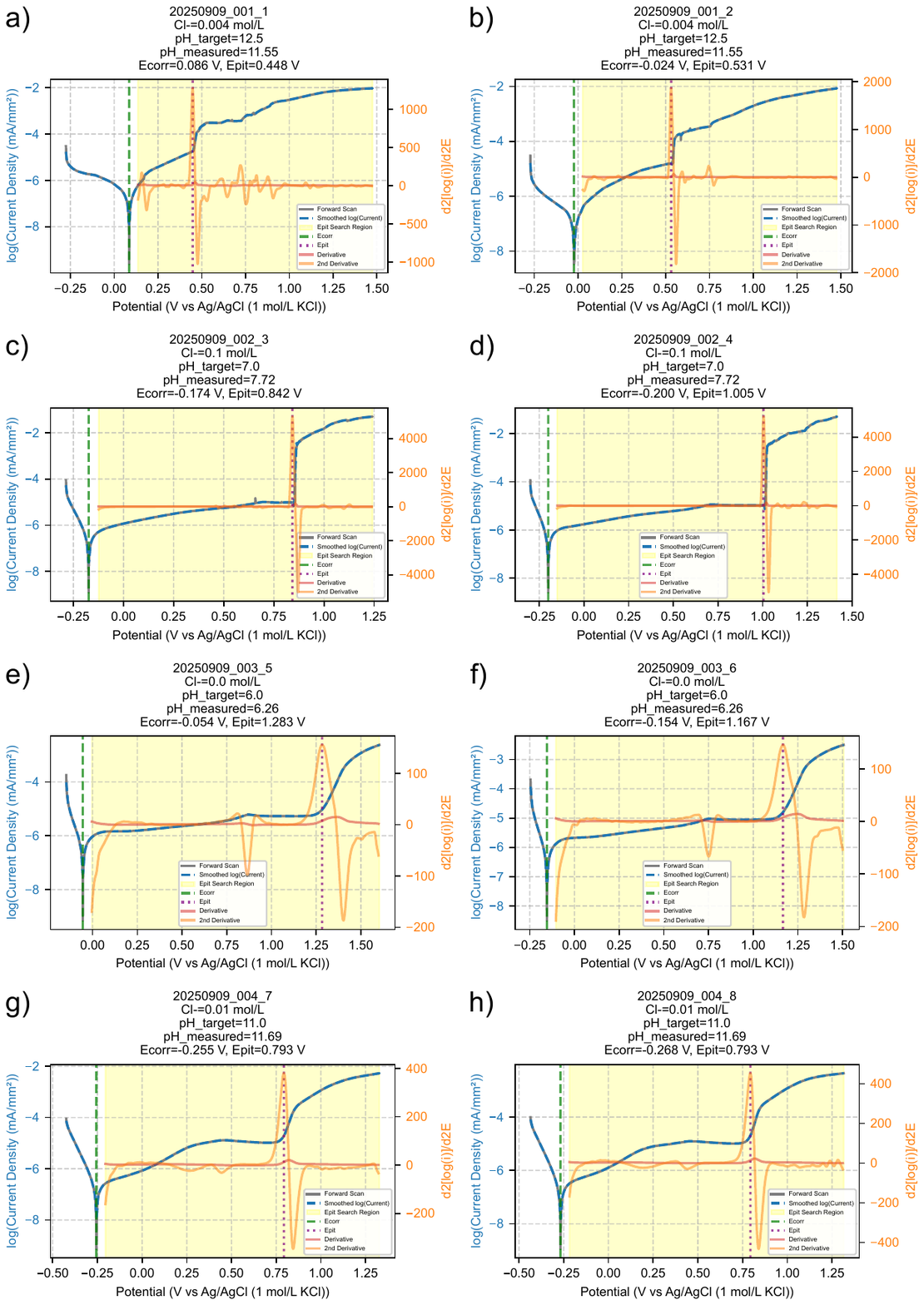}
    \caption{a) cell 1, b) cell 2, c) cell 3, d) cell 4, e) cell 5, f) cell 6, g) cell 7, h) cell 8. Each plot shows the logarithm of the current density vs. potential. The solid grey line is the raw data, blue dashed line is the smoothed data, solid red line is the smoothed data derivative, solid orange line is the smoothed data second derivative, dashed green line is the extracted $E_{\mathrm{corr}}$, and dotted purple line is the extracted $E_{\mathrm{pit}}$.}
    \label{si:autonomousStabilityDiagramScans4}
\end{figure}

\newpage
\begin{figure}[H]
    \centering
    \textbf{Potentodynamic Polarization scans for iteration 5 of stability diagram building}\par\medskip
    \includegraphics[width=0.80\textwidth]{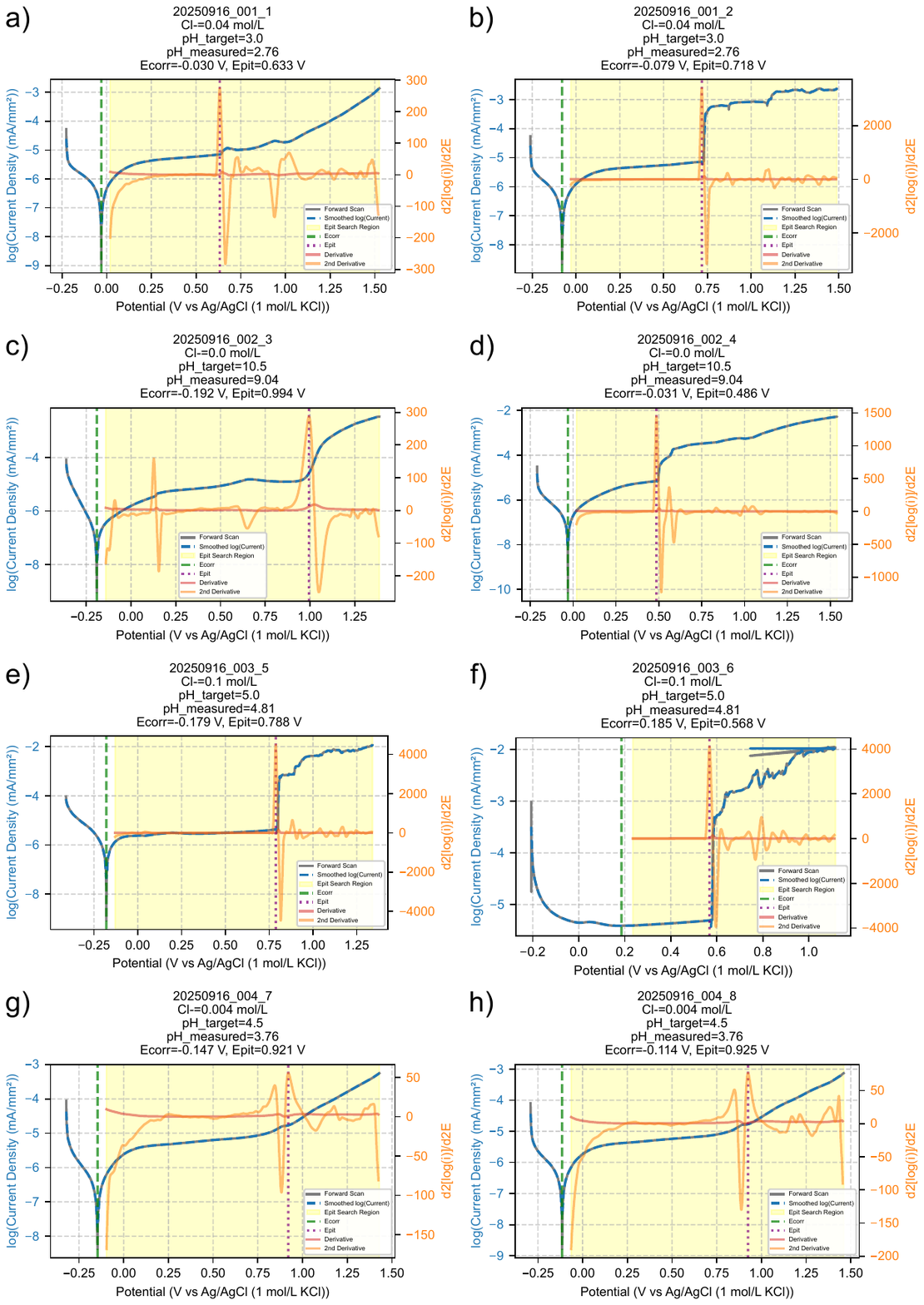}
    \caption{a) cell 1, b) cell 2, c) cell 3, d) cell 4, e) cell 5, f) cell 6, g) cell 7, h) cell 8. Each plot shows the logarithm of the current density vs. potential. The solid grey line is the raw data, blue dashed line is the smoothed data, solid red line is the smoothed data derivative, solid orange line is the smoothed data second derivative, dashed green line is the extracted $E_{\mathrm{corr}}$, and dotted purple line is the extracted $E_{\mathrm{pit}}$.}
    \label{si:autonomousStabilityDiagramScans5}
\end{figure}

\newpage
\begin{figure}[H]
    \centering
    \textbf{Potentodynamic Polarization scans for iteration 6 of stability diagram building}\par\medskip
    \includegraphics[width=0.80\textwidth]{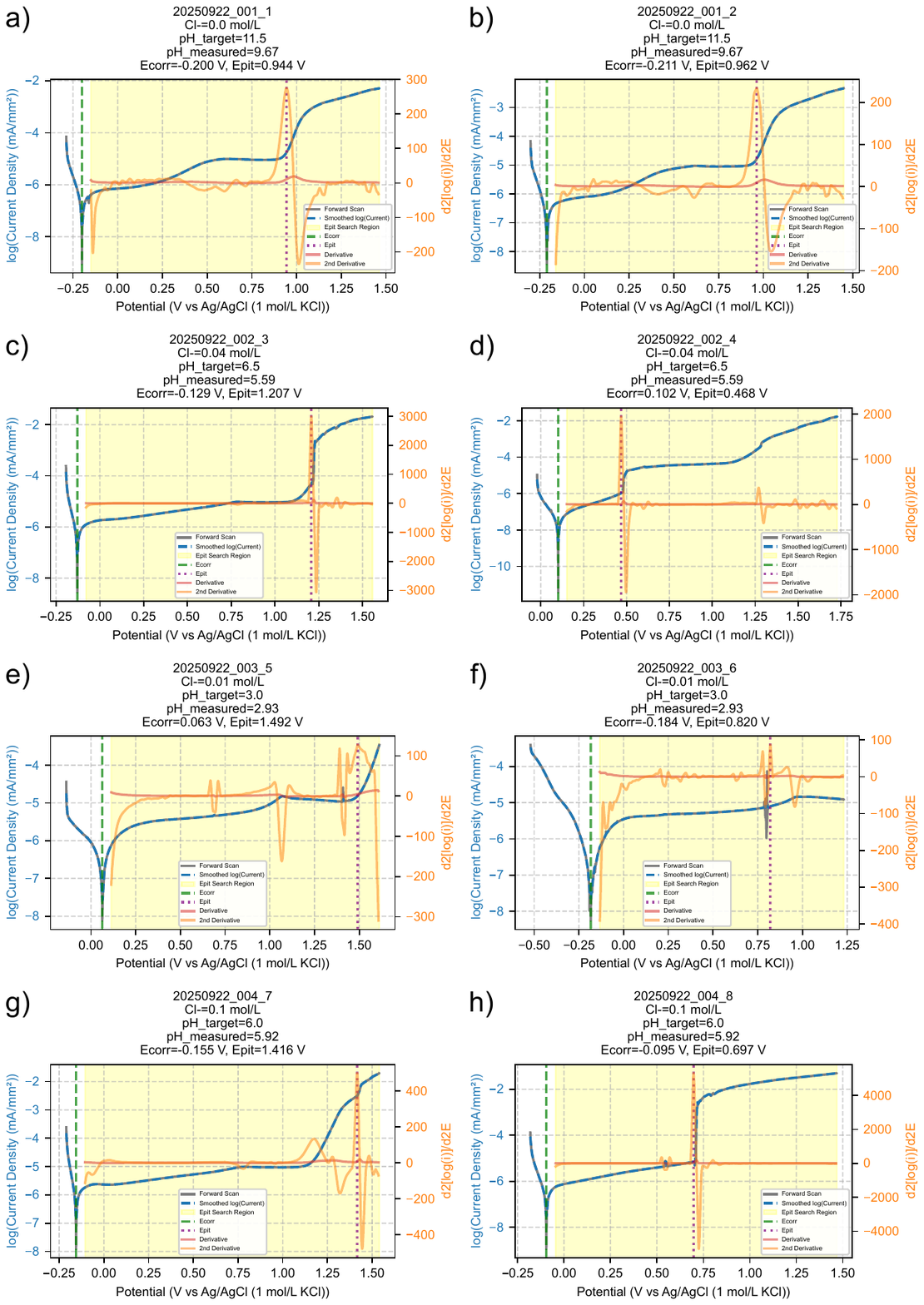}
    \caption{a) cell 1, b) cell 2, c) cell 3, d) cell 4, e) cell 5, f) cell 6, g) cell 7, h) cell 8. Each plot shows the logarithm of the current density vs. potential. The solid grey line is the raw data, blue dashed line is the smoothed data, solid red line is the smoothed data derivative, solid orange line is the smoothed data second derivative, dashed green line is the extracted $E_{\mathrm{corr}}$, and dotted purple line is the extracted $E_{\mathrm{pit}}$.}
    \label{si:autonomousStabilityDiagramScans6}
\end{figure}

\newpage
\section{Autonomous Stability Diagram Chloride dependence}
\label{sm:autonomousStabilityDiagramAdditionalAnalysis}
Below is all the $E_{\mathrm{pit}}$ values plotted against pH for the 56 PP scans performed as part of the autonomous stability diagram experiment, grouped by Cl$^-$. The traces corresponding to 0.04 mol/L and 0.1 mol/L Cl$^-$ show a positive correlation between pH and $E_{\mathrm{pit}}$, while the traces corresponding to 0 mol/L, 0.004 mol/L, and 0.01 mol/L Cl$^-$ show a negative correlation between pH and $E_{\mathrm{pit}}$. 

\begin{figure}[H]
    \centering
    \textbf{Measured pH vs $E_{\mathrm{pit}}$ for different Cl$^-$ concentrations}\par\medskip
    \includegraphics[width=\textwidth]{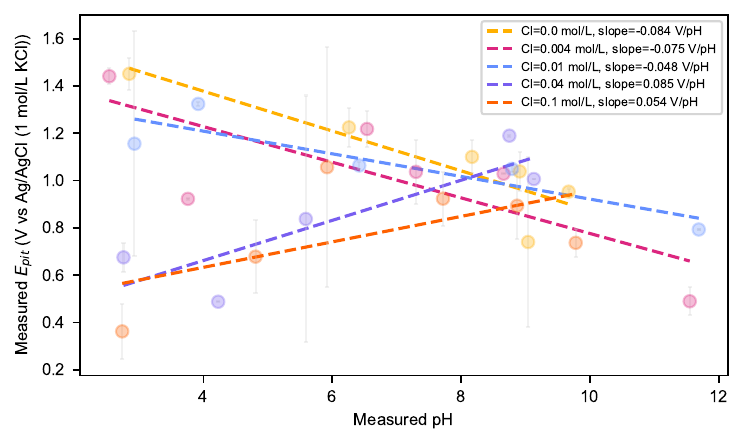}
    \caption{Scatter plot of measured pH vs. $E_{\mathrm{pit}}$ for the 56 PP scans performed as part of the autonomous stability diagram experiment, grouped by Cl$^-$ concentration, with 0 mol/L in yellow, 0.004 mol/L in pink, 0.01 mol/L in blue, 0.04 mol/L in purple, and 0.1 mol/L in orange.}
\end{figure}

\end{document}